\documentclass{jfm}
\usepackage{relsize}
\usepackage{caption}
\usepackage{subcaption}
\usepackage{graphicx}
\usepackage{natbib}
\usepackage{mathrsfs}
\usepackage{setspace}

\usepackage{amsmath,mathtools}
\usepackage[usenames,dvipsnames]{color} 
\usepackage{ifthen}

\usepackage{tabularx}
\usepackage{longtable}
\usepackage{listings}
\usepackage{booktabs}
\usepackage{float}
\usepackage{makecell}
\usepackage{epstopdf}
\usepackage{pifont} 
\usepackage{wasysym}
\usepackage{tikz}
\usepackage{upgreek}

\usepackage[normalem]{ulem}

\usepackage{flafter}  
\usepackage{ulem}

\newboolean{showcomments}
\setboolean{showcomments}{true}

\newcommand{\dennice}[1]{  \ifthenelse{\boolean{showcomments}}
{\textcolor{red}{(Dennice says:  #1)}}{}}

\def\mline{\vrule width4pt height2.5pt depth -2pt}
\def\dashed{\mline\hskip3.5pt\mline\thinspace}

\newtheorem{thm}{Theorem}
\newtheorem{pro}[thm]{Proposition} 
\newtheorem{definition}[thm]{Definition} 

\newtheorem{corollary}[thm]{Corollary}
\newenvironment{myproof}{\paragraph{\textbf{Proof:}}}{\hfill$\square$}

\usepackage{lmodern,pdftexcmds,amsmath}

\DeclareMathAlphabet{\mathsfbi}{OT1}{\sfdefault}{bx}{sl}
\DeclareMathVersion{sfletters}
\SetSymbolFont{letters}{sfletters}{OML}{ntxsfmi}{b}{it}

\makeatletter
\newcommand{\mathbfsbilow}[1]{%
  \text{\mathversion{sfletters}$\m@th#1$}%
}
\makeatother

\usepackage{etoolbox}

\newcommand{\rev}[1]{{\color{black}#1}} 

\newtoggle{thesis}
\togglefalse{thesis}

\begin{document}

\shorttitle{Structured input--output analysis of transitional channels} 
\shortauthor{C. Liu, D. F. Gayme} 

\title{Structured input--output analysis of transitional wall-bounded flows}

\author
 {
 Chang Liu\aff{1}
  \corresp{\email{changliu@jhu.edu}}, Dennice F. Gayme \aff{1}
  }

\affiliation
{
\aff{1}
Department of Mechanical Engineering, Johns Hopkins University, Baltimore, MD 21218, USA

}

\maketitle

\begin{abstract}

Input--output analysis of transitional channel flows has proven to be a valuable analytical tool for identifying important flow structures and energetic motions. The traditional approach abstracts the nonlinear terms as forcing that is unstructured, in the sense that \rev{this forcing} is not directly tied to the underlying nonlinearity in the dynamics. This paper instead employs a \emph{structured} singular value-based approach that preserves certain input--output properties of the nonlinear forcing function in an effort to recover the larger range of key flow features identified through nonlinear analysis, experiments, and direct numerical simulation (DNS) of transitional channel flows. Application of this method to transitional plane Couette and plane Poiseuille flows leads to \rev{not only the identification of} the streamwise coherent structures predicted through traditional input--output approaches,  \rev{ but also the} characterization of the oblique flow structures as those requiring the least energy to induce transition in agreement with DNS studies, and nonlinear optimal perturbation analysis. The proposed approach also captures the recently observed oblique turbulent bands that have been linked to transition in experiments and DNS with very large channel size. The ability to identify the larger amplification of the streamwise varying structures predicted from DNS and nonlinear analysis in both flow regimes suggests that the structured approach allows one to maintain the nonlinear effects associated with  \rev{weakening} of the lift-up mechanism, which is known to dominate the linear operator. Capturing this key nonlinear \rev{effect} enables the prediction of the wider range of known transitional flow structures within the analytical input--output modeling paradigm.


\end{abstract}

\section{Introduction}
\label{sec:introduction}

Interest in  transitional wall-bounded shear flow dates back to early studies by \citet{reynolds1883xxix}, who noted that the flow \rev{in a pipe} was sensitive to disturbances. Though much progress has been made, a full understanding of the phenomena has yet to be realized. One of the main challenges lies in the fact that linear stability analysis fails to accurately predict the Reynolds numbers at which flows are observed \rev{to} transition to turbulence. For example, plane Couette flow is linearly stable for any Reynolds number \citep{romanov1973stability} yet is observed to transition to turbulence at Reynolds numbers as low as $360\pm 10$ \citep{tillmark1992experiments}. This failure has led researchers to study the mechanisms underlying transition by instead analyzing energy growth. In particular, there has been an emphasis on characterizing the types of finite-amplitude perturbations that are most likely to lead to transition as well as, the flow structures that dominate in this regime, see e.g. \citet{schmid1992new,lundbladh1994threshold,reddy1998stability,philip2007scaling,duguet2010towards,Duguet2013,farano2015hairpin}.

\citet{reddy1998stability} examined the relative effect of different  transition-inducing flow perturbations in both plane Couette flow and Poiseuille flow through extensive direct numerical simulations (DNS). These authors observed that both streamwise vortices and oblique waves require less energy density than random noise to trigger transition \citep[figures 19 and 23]{reddy1998stability} in both flows. They further showed that in Poiseuille flow even perturbations in the form of Tollmien–Schlichting (TS) waves, which are linearly unstable at $Re>5772$ \citep{orszag1971accurate},  require larger energy density to trigger transition than either streamwise vortices or oblique waves \citep[figure 19]{reddy1998stability}. Similar behavior has been observed in studies of the transient energy growth and input--output response of the linearized Navier-Stokes (LNS) equations \citep{reddy1993energy,Jovanovic2005}. In fact, input--output analysis of channel flow suggests that streamwise constant structures have larger   energy growth than the linearly unstable TS waves, even at supercritical Reynolds numbers (i.e. above the Reynolds number at which the laminar flow is no longer linearly stable) \citep{jovanovic2004unstable,jovanovic2004modeling}. Studies of the LNS have indicated that streamwise vortical structures represent both the initial condition (optimal perturbation) that leads to the largest energy growth \citep{gustavsson1991energy,butler1992three,reddy1993energy,schmid2012stability}, as well as the type of structures that sustains the highest energy growth, see e.g. \citep{Farrell1993,Bamieh2001, Jovanovic2005}. The importance of streamwise vortices was also confirmed by \citet{bottin1998experimental}, who connected experimental results with this form of exact coherent structures in plane Couette flow.

On the other hand, the simulations of \citet{schmid1992new} and \citet{reddy1998stability}, as well as the experiments of  \cite{elofsson1998experimental} indicate that perturbations of oblique waves require slightly less energy than streamwise vortices to initiate transition.  Nonlinear optimal perturbations (NLOP) to plane Couette flow, i.e. the initial perturbations that require the least energy to transition the flow from  laminar to turbulent, also take the form of oblique waves that are localized in the streamwise direction, see e.g., \citep{duguet2010towards,Duguet2013,Monokrousos2011,rabin2012triggering,cherubini2013nonlinear,cherubini2015minimal}. For plane Poiseuille flow, hairpin vortices associated with the very short timescale of the Orr mechanism represent the NLOP \citep{farano2015hairpin,farano2016subcritical}. These results suggest that traditional linear analysis does not capture the full range of highly amplified structures in transitional flows.

Recent experiments and DNS of plane Couette flow with very large channel size ($\sim O(100)$ times the channel half-height) have also uncovered oblique turbulent bands (turbulent stripes) in the transitional flow regime of wall-bounded shear flows, see e.g. \citep{prigent2002large,prigent2003long, duguet2010formation,de2020transient,tuckerman2020patterns}. These turbulent-laminar patterns were also observed \rev{in DNS of transitional plane Poiseuille flow with sufficiently large channel size} \citep{tsukahara2005dns,xiong2015turbulent,tao2018extended,kanazawa2018lifetime,shimizu2019bifurcations,xiao2020growth,song2020trigger}. The presence of such structures was later confirmed by experiments \citep[figure 1]{tsukahara2014experimental,paranjape2019onset,paranjape2020oblique}. There is strong evidence that the mechanisms leading to the growth and maintenance of these oblique turbulent bands are nonlinear in both plane Couette \citep{barkley2007mean,tuckerman2011patterns,duguet2013oblique} and plane Poiseuille flow \citep{tuckerman2014turbulent}. That view has been further supported by analysis of 
exact equilibrium and traveling wave solutions of the nonlinear Navier-Stokes (NS) equations; see e.g., for plane Couette flow \citep{deguchi2015asymptotic,reetz2019exact} and Poiseuille flow \citep{paranjape2020oblique}.

The literature described above points to the benefit of nonlinear methods in characterizing the full range of flow structures in transitional channel flow. However, these methods have far larger computational costs than linear analysis methods; see e.g., \citep{Kerswell2014,Kerswell2018}. This trade-off between obtaining a more comprehensive characterization of the phenomena and analysis that is computationally tractable is long-standing. However, there is significant evidence to suggest that insight can be gained through parametrizing or bounding the effect of the nonlinearity rather than undertaking the full computational burden of resolving it. For example, \citet{kreiss1994bounds} and \citet{chapman2002subcritical} employed a bound on the nonlinearity to derive a finite-amplitude permissible perturbation that a flow could sustain while remaining laminar. More recently, finite-amplitude stability analysis of transitional shear flows employed local componentwise (sector) bounds on the nonlinearity and exploited the passivity of the nonlinear operator to develop linear matrix inequality (LMI) based approaches to compute bounds on permissible perturbations for a range of shear flow models; see e.g., \citet{kalur2020nonlinear,kalur2020stability,kalur2021estimating,liu2020input}. The inclusion of information about the nonlinear behavior produced results that matched simulation data better than those derived through previous linear approaches. Data-driven methods to parametrize or color (in space or time)   \rev{input (forcing) applied to the dynamics linearized around the turbulent mean velocity} have also enabled nonlinear effects to be captured within the input--output framework; leading to better prediction of flow statistics~\citep{chevalier2006state,Zare2017,morra2021color,nogueira2021forcing}. The effect of the nonlinearity in the NS equations has also been incorporated directly into input--output and resolvent analysis through shaping or parametrizing the forcing, e.g. by including larger amplitude forcing in the near-wall region \citep{jovanovic2001modeling,Hpffner2005}. 

\begin{figure}


    \centering
    \includegraphics[width=0.6\textwidth]{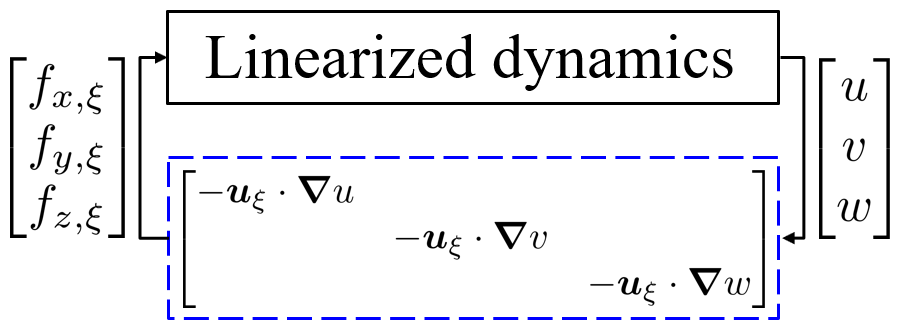}
    \caption{Block diagram representing  structured input--output feedback interconnection between the linearized dynamics and ({\color{blue}$\dashed$}) structured forcing (modeling the nonlinearity). In particular, each component of the forcing is modeled as an input--output mapping from the respective  component of velocity gradient $\boldsymbol{\nabla} u$, $\boldsymbol{\nabla} v$, $\boldsymbol{\nabla} w$ to the respective component forcings $f_{x,\xi}$, $f_{y,\xi}$, $f_{z,\xi}$ of the linearized dynamics with the gain $-\boldsymbol{u}_{\xi}$ defined in terms of the structured singular value of a linearized closed-loop system  response. }
    \label{fig:feedback_introduction}
\end{figure}

In this work, we build on this notion of including the effect of the nonlinearity within a computationally tractable linear framework using the concept of a \emph{structured} uncertainty, see e.g., \citep{packard1993complex,zhou1996robust}. In particular, we partition the NS equations \rev{into a feedback interconnection between the linearized dynamics and a model of the nonlinear forcing, as shown in figure \ref{fig:feedback_introduction}. We then structure the feedback to enforce a block-diagonal structure (bottom block outlined  by the blue dashed line {\color{blue}$\dashed$}). In particular, the}  feedback defines the componentwise inputs to the linearized momentum equations, which are modeled in terms of an uncertain gain $-\boldsymbol{u}_{\xi}$ of an input--output mapping from each component $\boldsymbol{\nabla} u$, $\boldsymbol{\nabla} v$ and $\boldsymbol{\nabla} w$ to the respective forcings $f_{x,\xi}$, $f_{y,\xi}$ and $f_{z,\xi}$.  We represent this gain using the \emph{structured} singular value \citep{doyle1982analysis,safonov1982stability},  $\mu$, which we \rev{use} to define the \rev{largest} gain under the structured forcing \citep{packard1993complex}. Conceptually the approach allows us to develop a feedback interconnection between the LNS and a structured forcing that is explicitly constrained to preserve the componentwise structure of the nonlinearity in the NS equations. 

 \emph{Structured} input--output analysis shares the advantages of all methods employing the spatio-temporal frequency response based analysis techniques upon which it is built, see e.g., \citep{Farrell1993, Bamieh2001,Jovanovic2005,McKeon2010,mckeon2013experimental,mckeon2017engine,illingworth2018estimating,vadarevu2019coherent,madhusudanan2019coherent,symon2021energy,liu2019vorticity,liu2019input}. Of greatest interest in this work is its computational tractability versus nonlinear approaches and the lack of finite channel size effects that can plague both DNS and experimental studies. This approach is most closely related to the analysis of the largest singular value ($\mathcal{H}_\infty$ norm) of the spatio-temporal frequency response of the linearized dynamics (top-block of figure \ref{fig:feedback_introduction}), which measures the structure that sustains the highest input--output growth, see e.g., \citet[chapter 8.1.2]{jovanovic2004modeling}; \citet{schmid2007nonmodal,hwang2010amplification,Hwang2010Linear,illingworth2020streamwise}. However, in that work, the forcing is assumed to excite the
dynamics at all frequencies (e.g., delta-correlated spatio-temporal white noise); in this sense, it can be thought of as the open-loop response of the top-block in figure \ref{fig:feedback_introduction}.

We apply the proposed \emph{structured} input--output analysis to transitional plane Couette and plane Poiseuille flow. The results indicate that the addition of a structured feedback interconnection enables identification and analysis of the wider range of transition-inducing flow structures identified in the literature without the computational burden of nonlinear optimization or extensive simulations. More specifically, the results for transitional plane Couette flow reproduce results from DNS based analysis \citep{reddy1998stability} and predictions of NLOP approaches \citep{rabin2012triggering}, which both indicate that oblique waves require less energy to \rev{induce transition} than the streamwise elongated structures emphasized in traditional input--output analysis. In plane Poiseuille flow, these transition-inducing flow structures are consistent with DNS \citep{reddy1998stability} emphasizing oblique waves and NLOP analysis that highlights the importance of spatially localized structures with streamwise wavelengths larger than their spanwise extent ~\citep{farano2015hairpin}. The proposed approach also reproduces the characteristic wavelengths and angle of the oblique turbulent band observed in very large channel studies of transitional plane Couette flow \citep{prigent2003long}. The wavelengths of oblique turbulent bands in transitional plane Poiseuille flow with very large channel size \citep{kanazawa2018lifetime} also fall within the range of flow structures showing large structured input--output response. 

The agreement between prediction\rev{s} from structured input--output analysis and observation in experiments, DNS, and NLOP show that this framework captures important nonlinear effects.
In particular, the results suggest that restricting the feedback in a componentwise manner preserves the structure of the nonlinear mechanisms that weaken the streaks developed through the lift-up effect, in which cross-stream forcing amplifies streamwise streaks \citep{ellingsen1975stability,landahl1975wave,Brandt2014}. \rev{Traditional input--output analysis instead predicts the dominance of streamwise elongated structures associated with the lift-up mechanism,} see e.g. the discussion in \citet{jovanovic2020bypass}. \rev{An examination of Reynolds number trends supports the notion that imposing a structured feedback interconnection based on certain input--output properties associated with the nonlinearity in the NS equations leads to a weakening of the amplification of streamwise elongated structures.} 

The remainder of this paper is organized as follows. Section \ref{sec:structured_IO} describes the flow configurations of interest and describes the details of the structured input--output analysis approach. Section \ref{sec:result} analyzes the results obtained from the application of structured input--output analysis to both plane Couette and plane \rev{Poiseuille} flow. \rev{We then analyze  Reynolds number dependence in \S\  \ref{sec:Re_dependence}.} This paper is concluded in \S\ \ref{sec:conclusion}.

\section{Formulating the structured input--output model }
\label{sec:structured_IO}

\subsection{Governing Equations}

\begin{figure}

	(a) \hspace{0.49\textwidth} (b)
	
\centering
\includegraphics[width=0.49\textwidth]{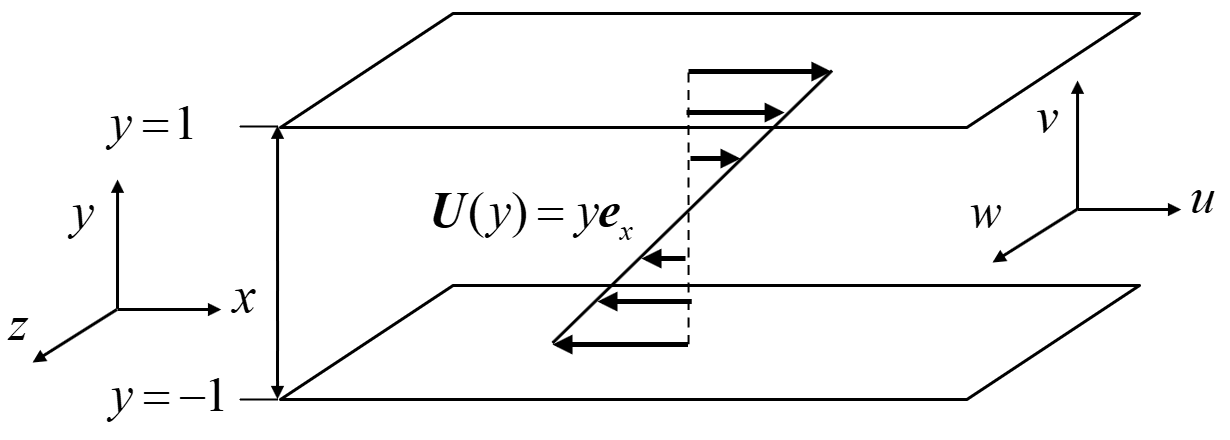}
\includegraphics[width=0.49\textwidth]{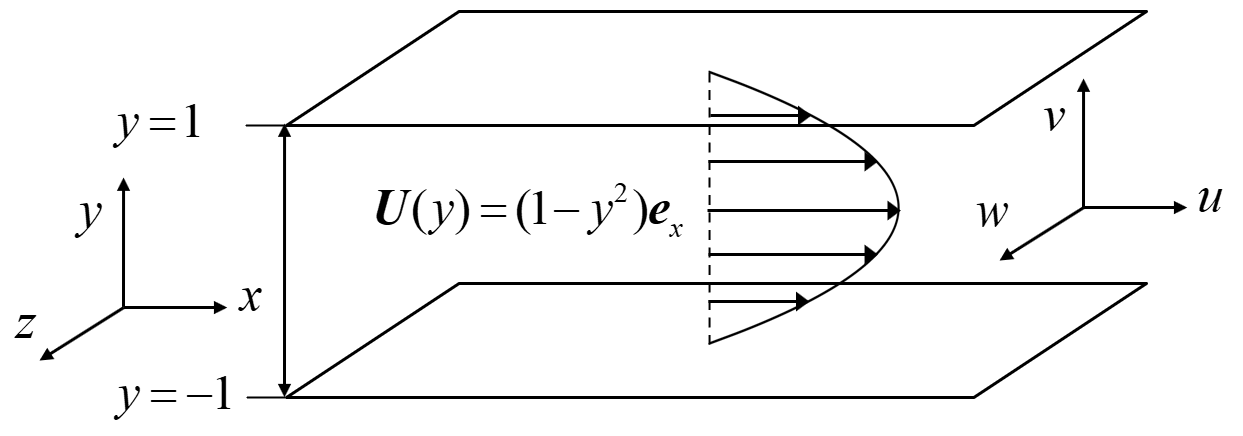}
    \caption{ Illustrations of flows between two parallel flat plates: (a) plane Couette flow with laminar base flow $\boldsymbol{U}(y)=y\boldsymbol{e}_x$ and (b) plane Poiseuille flow with laminar base flow $\boldsymbol{U}(y)=(1-y^2)\boldsymbol{e}_x$. }
    \label{fig:flow_config_cou_poi}
\end{figure}

We consider incompressible flow between two infinite parallel plates and employ $x$, $y$, and $z$ to respectively denote the streamwise, wall-normal, and spanwise directions. The corresponding velocity components are denoted \rev{by} $u$, $v$, and $w$. The coordinate frames and configurations \rev{being used} for plane Couette and plane Poiseuille flows are shown in figure \ref{fig:flow_config_cou_poi}.
We express the velocity field as a vector $\boldsymbol{u}_{\text{tot}}=\begin{bmatrix}u_{\text{tot}} & v_{\text{tot}} & w_{\text{tot}}\end{bmatrix}^\text{T}$ with $^\text{T}$ indicating the transpose. We then decompose the velocity field into the sum of a laminar base flow ($U(y)=y$ for plane Couette flow and $U(y)=1-y^2$ for plane Poiseuille flow) and fluctuations about the base flow $\boldsymbol{u}$; i.e., $\boldsymbol{u}_{\text{tot}}=U(y)\boldsymbol{e}_x+\boldsymbol{u}$ with $\boldsymbol{e}_x$ denoting the $x$-direction (streamwise) unit vector. The pressure field is \rev{similarly decomposed  into} $p_{\text{tot}}=P+p$. The dynamics of the fluctuations $\boldsymbol{u}$ and $p$ are governed by the \rev{NS} equations:
\begin{subequations} \label{eq:NS_All}
\begin{align}
\partial_{t} \boldsymbol{u}  
+  U\partial_x \boldsymbol{u}  + v\,\frac{dU }{dy}\boldsymbol{e}_x +\boldsymbol{\nabla} p
-\frac{1}{Re}{\nabla}^2 \boldsymbol{u}
 &=- \boldsymbol{u} \cdot \boldsymbol{\nabla} \boldsymbol{u}, \label{eq:NSDecompf1} \\
\boldsymbol{\nabla} \cdot \boldsymbol{u}&=0. \label{eq:NSDecompf2}
\end{align}
\end{subequations} 
Here, the spatial variables are normalized by the channel half-height $h$: e.g., $y=y_*/h\in[-1,1]$, where the subscript $*$ indicates dimensional quantities. The velocity is normalized by a nominal characteristic velocity $U_n$, where $\pm U_n$ is the velocity at \rev{the} channel walls for plane Couette flow, and $U_n$ is \rev{the} channel centerline velocity for plane Poiseuille flow. Time and pressure are normalized by $h/U_n$ and $\rho U_n^2$, respectively. The Reynolds number is defined as $Re=U_nh/\nu$, where $\nu$ is the kinematic viscosity. In equation \eqref{eq:NS_All},  $\boldsymbol{\nabla}:=\begin{bmatrix}\partial_x & \partial_y & \partial_z\end{bmatrix}^\text{T}$ represents the gradient operator, and $\nabla^2:=\partial_x^2+\partial_y^2+\partial_z^2$ represents the Laplacian operator. We impose no-slip boundary conditions at the wall; i.e., $\boldsymbol{u}(y=\pm 1)=\boldsymbol{0}$ for both flows. Finally, we write the nonlinear term in equation \eqref{eq:NSDecompf1} as
\begin{equation}
\boldsymbol{f}:=-\boldsymbol{u}\cdot \boldsymbol{\nabla }\boldsymbol{u}=\begin{bmatrix}-\boldsymbol{u}\cdot \boldsymbol{\nabla }u &-\boldsymbol{u}\cdot \boldsymbol{\nabla }v& -\boldsymbol{u}\cdot \boldsymbol{\nabla }w\end{bmatrix}^\text{T}=:\begin{bmatrix}f_x&f_y&f_z\end{bmatrix}^\text{T}, 
     \label{eq:f_nonlinear}
\end{equation}
where $=:$ indicates that the right-hand side is defined by the left-hand side. We refer to $f_x$, $f_y$, and $f_z$ as the respective streamwise, wall-normal, and spanwise components of the nonlinearity and collectively as the nonlinear components of \eqref{eq:NS_All}. This expression of the nonlinearity as forcing terms makes \eqref{eq:NS_All} into a set of forced linear evolution equations. \rev{This approach builds on the growing body of work that has shown promise in capturing critical features of this forced system response using linear analysis techniques, see e.g. the reviews of  \citet{schmid2007nonmodal,mckeon2017engine,jovanovic2020bypass} and the references therein. }

We next construct the model of the nonlinearity that will allow us to build the feedback interconnection of figure \ref{fig:feedback_introduction}. The velocity field $-\boldsymbol{u}$ in \eqref{eq:f_nonlinear}  associated with the forcing components can be viewed as the gain operator of an input--output system in which the velocity gradients  $\boldsymbol{\nabla} u$, $\boldsymbol{\nabla} v$, $\boldsymbol{\nabla} w$ act as the respective inputs and the forcing components $f_x$, $f_y$ and $f_z$ act as the respective output. It is this gain that we seek to model through $-\boldsymbol{u}_{\xi}$ in figure \ref{fig:feedback_introduction}. This input--output model of the  nonlinear components is given by
\begin{equation}
    \boldsymbol{f}_{\xi}:=-\boldsymbol{u}_{\xi}\cdot \boldsymbol{\nabla }\boldsymbol{u}=\begin{bmatrix}-\boldsymbol{u}_{\xi}\cdot \boldsymbol{\nabla }u &-\boldsymbol{u}_{\xi}\cdot \boldsymbol{\nabla }v& -\boldsymbol{u}_{\xi}\cdot \boldsymbol{\nabla }w\end{bmatrix}^\text{T}=:\begin{bmatrix}f_{x,\xi}&f_{y,\xi}&f_{z,\xi}\end{bmatrix}^\text{T}, 
     \label{eq:f_uncertain_model}
\end{equation}
where \rev{$-\boldsymbol{u}_{\xi}=[-u_\xi,-v_\xi,-w_\xi]^\text{T}$} maps the corresponding velocity gradient into each component of the modeled  forcing driving linearized dynamics.  The next subsection describes how we construct \rev{this} input--output map so that it enables us to analyze the perturbations that are most likely to induce transition using the structured singular value formalism \citep{packard1993complex,zhou1996robust}.

\subsection{Structured input--output response}
\label{subsec:structured_uncertainty}

We now define the spatio-temporal frequency response $\mathcal{H_\nabla}(y;k_x,k_z,\omega)$ that will form the basis of \rev{the} structured input--output response. We first perform the standard transformation to express the dynamics \eqref{eq:NS_All} in terms of the wall-normal velocity $v$ and wall-normal vorticity $\omega_y:=\partial_z u-\partial_x w$ \citep{schmid2012stability}, which enforces \eqref{eq:NSDecompf2} and eliminates the pressure dependence. \rev{This formulation similarly imposes the divergence-free condition on the forcing model, since  any component of the input forcing that can be written as the gradient of a scalar function $\widehat{\boldsymbol{f}}_{\phi}=\widehat{\boldsymbol{\nabla}}\widehat{\phi}$ will be absorbed into the pressure gradient and eliminated.} We then exploit the shift-invariance in the $(x,z)$ spatial directions of the two flow configurations of interest and assume invariance to shifts in $t$, which allows us to perform the following triple Fourier transform for a variable $\psi$:
\begin{equation}
\widehat{\psi}(y;k_x,k_z,\omega):=\int\limits_{-\infty}^{\infty} \int\limits_{-\infty}^{\infty}\int\limits_{-\infty}^{\infty}\psi(x,y,z,t)e^{-\text{i}(k_x x + k_z z+\omega t )}\rev{\,dx\,dz\,dt}, \label{eq:def_fourier_xz}
\end{equation}
where $\text{i}=\sqrt{-1}$ is the imaginary unit and $\omega$ is the temporal frequency.  $k_x = 2\pi/\lambda_x$ and $k_z = 2\pi/\lambda_z$ are the respective dimensionless $x$ and $z$ wavenumbers.

The resulting system of equations describing the transformed linearized equations subject to the modeled forcing $\boldsymbol{f}_\xi$ is 
\begin{subequations}
\label{eq:ABC_frequency}
\begin{equation}
    \text{i}\omega\begin{bmatrix}
    \widehat{v}\\
    \widehat{\omega}_y
    \end{bmatrix}=\widehat{\mathcal{A}}\begin{bmatrix}
    \widehat{v}\\
    \widehat{\omega}_y
    \end{bmatrix}+\widehat{\mathcal{B}}\begin{bmatrix}\widehat{f}_{x,\xi}\\
    \widehat{f}_{y,\xi}\\
    \widehat{f}_{z,\xi}\end{bmatrix},\;\;
    \begin{bmatrix}\widehat{u}\\
    \widehat{v}\\
    \widehat{w}\end{bmatrix}=\widehat{\mathcal{C}}\begin{bmatrix}
    \widehat{v}\\
    \widehat{\omega}_y
    \end{bmatrix}.\tag{\theequation a,b}
\end{equation}
\end{subequations}
The operators in equation \eqref{eq:ABC_frequency} are defined following \citet{Jovanovic2005}:
\begingroup
\allowdisplaybreaks
\begin{subequations}
\label{eq:operator_ABC}
\begin{align}
    \widehat{\mathcal{A}}(k_x,k_z):=&\begin{bmatrix}\widehat{{\nabla}}^2 & 0\\
    0 & \mathcal{I}
    \end{bmatrix}^{-1}\begin{bmatrix}
    -\text{i}k_xU{\widehat{\nabla}}^2+\text{i}k_xU''+\widehat{{\nabla}}^4/Re & 0\\
    -\text{i}k_zU' & -\text{i}k_x U+\widehat{{\nabla}}^2/Re
    \end{bmatrix},\label{eq:operator_ABC_A}\\
    \mathcal{\widehat{B}}(k_x,k_z):=& \begin{bmatrix}\widehat{{\nabla}}^2 & 0\\
    0 & \mathcal{I}
    \end{bmatrix}^{-1}
    \begin{bmatrix}
    -\text{i}k_x\partial_y & -(k_x^2+k_z^2) & -\text{i}k_z \partial_y\\
    \text{i}k_z & 0 & -\text{i}k_x
    \end{bmatrix},\label{eq:operator_ABC_B}\\
    \mathcal{\widehat{C}}(k_x,k_z):=&\frac{1}{k_x^2+k_z^2}\begin{bmatrix}
    \text{i}k_x\partial_y & -\text{i}k_z\\
    k_x^2+k_z^2 & 0\\
    \text{i}k_z \partial_y & \text{i}k_x
    \end{bmatrix},\label{eq:operator_ABC_C}
\end{align}
\end{subequations}
\endgroup
where $U':=dU(y)/dy$, $U'':=d^2U(y)/dy^2$, $\widehat{{\nabla}}^2:=\partial_{yy}-k_x^2-k_z^2$, and $\widehat{\nabla}^4:=\partial^{(4)}_{y}-2(k_x^2+k_z^2)\partial_{yy}+(k_x^2+k_z^2)^2$. The boundary conditions, which can be derived from the no-slip conditions, are
\begin{equation}
    \widehat{v}(y=\pm 1)=
    \frac{\partial \widehat{v}}{\partial y}(y=\pm1)=\widehat{\omega}_y(y=\pm 1)=0.
    \label{eq:BC}
\end{equation}

The spatio-temporal frequency response $\mathcal{H}$ of the system in  \eqref{eq:ABC_frequency} that maps the input forcing $\boldsymbol{\widehat{f}}_{\xi}(y;k_x,k_z,\omega)$ to the velocity vector $\boldsymbol{\widehat{u}}(y;k_x,k_z,\omega)$  at the same spatial-temporal wavenumber-frequency triplet; i.e., $\boldsymbol{\widehat{u}}(y;k_x,k_z,\omega)=\mathcal{H}(y;k_x,k_z,\omega)\boldsymbol{\widehat{f}}_{\xi}(y;k_x,k_z,\omega)$ is given by 
\begin{equation}
    \mathcal{H}(y;k_x,k_z,\omega):=\widehat{\mathcal{C}}\left(\text{i}\omega\,\mathcal{I}_{2\times 2}-\widehat{\mathcal{A}}\right)^{-1}\widehat{\mathcal{B}}.
     \label{eq:linearized_ABC}
\end{equation}
Here $\mathcal{I}_{2\times 2}:=\text{diag}(\mathcal{I},\mathcal{I})$, where $\mathcal{I}$ is the identity operator and $\text{diag}(\cdot)$ indicates a block diagonal operation. Following the language in \citet{jovanovic2020bypass}, we also refer to $\mathcal{H}(y; k_x,k_y, \omega)$ defined in \eqref{eq:linearized_ABC} as the frequency response operator.

\begin{figure}

	(a) \hspace{0.7\textwidth} (b)
	
    \centering
    \includegraphics[width=0.7\textwidth]{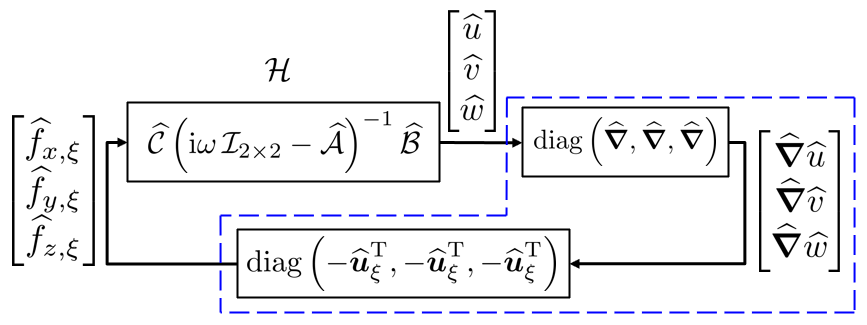}
    \hspace{0.2in}
    \includegraphics[width=0.8in,trim=-0 -0.5in 0 0]{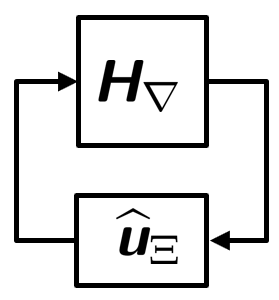}
    \caption{  Illustration of structured input--output analysis: (a) a componentwise description,  where blocks inside of ({\color{blue}$\dashed$}, blue) represent the modeled forcing in equation \eqref{eq:feedback_structured_uncertainty} corresponding to the bottom block of figure \ref{fig:feedback_introduction}; (b) a high-level description after discretization.}
    \label{fig:feedback_detail}
\end{figure}

The linear form of \eqref{eq:f_uncertain_model}  allows us to also  perform the spatio-temporal Fourier transform \eqref{eq:def_fourier_xz} on  \rev{this forcing} model to obtain 
\begin{equation}
\label{eq:feedback_structured_uncertainty_1}
\boldsymbol{\widehat{f}}_{\xi}=-\boldsymbol{\widehat{u}}_{\xi}\cdot\boldsymbol{\widehat{\nabla}} \boldsymbol{\widehat{u}}=
\begin{bmatrix}
    \widehat{f}_{x,\xi}\\
    \widehat{f}_{y,\xi}\\
    \widehat{f}_{z,\xi}
    \end{bmatrix}=\begin{bmatrix}
    -\boldsymbol{\widehat{u}}_{\xi}^{\text{T}} & &\\
     & -\boldsymbol{\widehat{u}}_{\xi}^{\text{T}} & \\
      & & -\boldsymbol{\widehat{u}}_{\xi}^{\text{T}}
    \end{bmatrix}\begin{bmatrix}
\boldsymbol{\widehat{\nabla}}\widehat{u}\\
\boldsymbol{\widehat{\nabla}}\widehat{v}\\
\boldsymbol{\widehat{\nabla}}\widehat{w}
\end{bmatrix},
\end{equation}
which can be decomposed as
\begin{equation}
\begin{bmatrix}
    \widehat{f}_{x,\xi}\\
    \widehat{f}_{y,\xi}\\
    \widehat{f}_{z,\xi}
    \end{bmatrix}=\text{diag}\left(-\widehat{\boldsymbol{u}}_{\xi}^{\text{T}} ,-\widehat{\boldsymbol{u}}_{\xi}^{\text{T}},-\widehat{\boldsymbol{u}}_{\xi}^{\text{T}} \right)\text{diag}\left(\boldsymbol{\widehat{\nabla}},\boldsymbol{\widehat{\nabla}},\boldsymbol{\widehat{\nabla}}\right)\begin{bmatrix}
\widehat{u}\\
\widehat{v}\\
\widehat{w}
\end{bmatrix}.
\label{eq:feedback_structured_uncertainty}
\end{equation}
This decomposition of the forcing function is illustrated in the two blocks inside the blue dashed line ({\color{blue}$\dashed$}) in figure \ref{fig:feedback_detail}(a), where the velocity field arising from the spatio-temporal frequency response $\mathcal{H}$ is the input and the forcing is the output. 

In order to isolate the gain $-\boldsymbol{u}_{\xi}$ that we seek to model, it is analytically convenient to combine the linear gradient operator with the spatio-temporal frequency response. We denote the resulting  modified frequency response operator as
\begin{equation}
     \mathcal{H}_{\nabla}(y;k_x,k_z,\omega):=\text{diag}\left(\boldsymbol{\widehat{\nabla}},\boldsymbol{\widehat{\nabla}},\boldsymbol{\widehat{\nabla}}\right)\mathcal{H}(y;k_x,k_z, \omega).
    \label{eq:H_operator_grad}
\end{equation}
\rev{We note that this operator in \eqref{eq:H_operator_grad} can be also obtained by modifying $\widehat{\mathcal{C}}$ in equation \eqref{eq:linearized_ABC} such that the output corresponds to a vectorized velocity gradient. Then, we} redraw the system as a feedback interconnection between this linear operator in \eqref{eq:H_operator_grad} and the structured uncertainty  
\begin{equation}
    \widehat{\boldsymbol{u}}_{\Upxi}:=\text{diag}\left(-\widehat{\boldsymbol{u}}_{\xi}^{\text{T}} ,-\widehat{\boldsymbol{u}}_{\xi}^{\text{T}},-\widehat{\boldsymbol{u}}_{\xi}^{\text{T}} \right).
    \label{eq:Delta_block_diag}
\end{equation}
The structured uncertainty $\widehat{\boldsymbol{u}}_{\Upxi}$ in \eqref{eq:Delta_block_diag} has a block-diagonal structure such that the resulting feedback interconnection leads to a forcing model that retains the componentwise structure of the nonlinearity. Figure \ref{fig:feedback_detail}(b) describes the resulting feedback interconnection between the modified spatio-temporal frequency response  and the structured uncertainty, where $\mathsfbi{H}_{\nabla}$ and $\mathbfsbilow{\widehat{u}}_{\Upxi}$ respectively represent the spatial discretizations (numerical approximations) of  $\mathcal{H}_{\nabla}$ in \eqref{eq:H_operator_grad} and $\boldsymbol{\widehat{u}}_{\Upxi}$ in \eqref{eq:Delta_block_diag}.

We are interested in characterizing the perturbations \rev{ associated with the most amplified flow structures under structured forcing.} This amplification under structured forcing can be quantified by the structured singular value of the modified frequency response operator $\mathcal{H}_\nabla$; see e.g., \citet[definition 3.1]{packard1993complex}; \citet[definition 11.1]{zhou1996robust}, which is defined as follows.

\begin{definition}\label{def:mu} Given wavenumber and frequency pair $(k_x,k_z,\omega)$, the structured singular value $\mu_{\mathbfsbilow{\widehat{U}}_{\Upxi}}\left[\mathbfsbilow{H}_{\nabla}(k_x,k_z,\omega)\right]$ is defined \rev{as}
\begin{align}
    \mu_{\mathbfsbilow{\widehat{U}}_{\Upxi}}\left[\mathbfsbilow{H}_{\nabla}(k_x,k_z,\omega)\right]:=\frac{1}{\text{min}\{\bar{\sigma}[\mathbfsbilow{\widehat{u}}_{\Upxi}]\,:\,\mathbfsbilow{\widehat{u}}_{\Upxi}\in \mathbfsbilow{\widehat{U}}_{\Upxi},\,\text{det}[\mathsfbi{I}-\mathbfsbilow{H}_{\nabla}(k_x,k_z,\omega)\mathbfsbilow{\widehat{u}}_{\Upxi}]=0\}},\label{eq:mu}
\end{align}
unless no $\mathbfsbilow{\widehat{u}}_{\Upxi}\in \mathbfsbilow{\widehat{U}}_{\Upxi}$ makes $\mathsfbi{I}-\mathbfsbilow{H}_{\nabla}\mathbfsbilow{\widehat{u}}_{\Upxi}$ singular, in which case $\mu_{\mathbfsbilow{\widehat{U}}_{\Upxi}}[\mathbfsbilow{H}_{\nabla}]:=0$.

Here, $\bar{\sigma}[\cdot]$ is the largest singular value, $\text{det}[\cdot]$ is the determinant of the argument, and $\mathsfbi{I}$ is the identity matrix. The subscript of $\mu$ in \eqref{eq:mu} is a set $\mathbfsbilow{\widehat{U}}_{\Upxi}$ containing all uncertainties having the same block-diagonal structure as $\mathbfsbilow{\widehat{u}}_{\Upxi}$; i.e.,
\begin{align}
\mathbfsbilow{\widehat{U}}_{\Upxi}:=\left\{\text{diag}\left(-\mathbfsbilow{\widehat{u}}_{\xi}^{\text{T}},-\mathbfsbilow{\widehat{u}}_{\xi}^{\text{T}},-\mathbfsbilow{\widehat{u}}_{\xi}^{\text{T}}\right):-\mathbfsbilow{\widehat{u}}^{\text{T}}_{\xi}\in \mathbb{C}^{N_y\times 3N_y}\right\},
\label{eq:uncertain_set}
\end{align}
where $N_y$ denotes the number of grid points in $y$.
\end{definition}

\rev{For ease of computation and analysis, the form of the structured uncertainty in equation (\ref{eq:uncertain_set}) allows the full degrees of freedom for the complex matrix $ -\mathbfsbilow{\widehat{u}}^{\text{T}}_{\xi}\in \mathbb{C}^{N_y\times 3N_y}$. A natural refinement to better represent the physics would be to enforce a diagonal structure for each of the sub-blocks of this matrix. This approach is not pursued here because it requires extensions of both the analysis and computational tools to properly evaluate the response. These extensions are beyond the scope of the current work.}

 \rev{The largest structured singular value across all temporal frequencies characterizes the largest response associated with a stable structured feedback interconnection (i.e. the full block diagram in figure \ref{fig:feedback_detail}(b)). Here, stability is defined in terms of} the small gain theorem \citep[theorem 11.8]{zhou1996robust}.
\begin{pro} [Small Gain Theorem]
Given $0<\beta<\infty$ and wavenumber pair $(k_x,k_z)$. The loop shown in figure \ref{fig:feedback_detail}(b) is stable for all $\mathbfsbilow{\widehat{u}}_{\Upxi}\in\mathbfsbilow{\widehat{U}}_{\Upxi} $ with $\|\mathbfsbilow{\widehat{u}}_{\Upxi}\|_{\infty}:=\underset{\omega \in \mathbb{R}}{\text{sup}}\,\bar{\sigma}[\mathbfsbilow{\widehat{u}}_{\Upxi}]<\frac{1}{\beta}$ if and only if:
\begin{align}
    \|\mathcal{H}_{\nabla}\|_{\mu}(k_x,k_z):=\underset{\omega \in \mathbb{R}}{\text{sup}}\,\mu_{\mathbfsbilow{\widehat{U}}_{\Upxi}}\left[\mathbfsbilow{H}_{\nabla}(k_x,k_z,\omega)\right]\leq \beta. 
    \label{eq:H_nabla_mu}
\end{align}
\label{pro:small_gain}
\end{pro}
\noindent Here, $\text{sup}$ represents supremum (least upper bound) operation, and we abuse the notation by writing $\|\cdot\|_{\mu}$ \citep{packard1993complex}, although $\mu$ is not a proper norm (i.e. it does not necessarily satisfy the triangle inequality).  This value $\|\mathcal{H}_{\nabla}\|_{\mu}(k_x,k_z)$ in \eqref{eq:H_nabla_mu} directly quantifies \rev{most amplified} flow structures \rev{(characterized by the associated ($k_x,k_z$) pair) under structured forcing.} \rev{This $\|\mathcal{H}_{\nabla}\|_{\mu}(k_x,k_z)$ is closely related to input--output analysis based on the $\mathcal{H}_{\infty}$ norm \citep{jovanovic2004modeling,schmid2007nonmodal,illingworth2020streamwise} and characterizations of transient growth (see e.g. \citep{schmid2007nonmodal}), where flow structures with high amplification under external input forcing or high transient energy growth are associated with transition.}

\rev{
 \subsection{Numerical Method}}
\label{subsec:numerical}

The operators in equation \eqref{eq:operator_ABC} are discretized using the Chebyshev differentiation matrices generated by the MATLAB routines of \citet{Weideman2000}. The boundary conditions in equation \eqref{eq:BC} are implemented following \citet[chapters 7 and 14]{trefethen2000spectral}. We employ the Clenshaw--Curtis quadrature \citep[chapter 12]{trefethen2000spectral} in computing both singular and structured singular values \rev{to ensure that they are} independent of the number of Chebyshev spaced wall-normal grid points. The numerical implementation of the operators is validated through comparisons of the plane Poiseuille flow results for computations of the  $\mathcal{H}_\infty$ norm in \citet[chapter 8.1.2]{jovanovic2004modeling} and \citet[figure 5]{schmid2007nonmodal}, and plane Couette flow in \citet[chapter 8.2]{jovanovic2004modeling}. We use $N_y=30$ collocation points (\rev{excluding the boundary points}), which is the same number employed in \citet{Jovanovic2005,jovanovic2004modeling}. We verified that doubling the number of collocation points in the wall-normal direction does not alter results, indicating grid convergence. We employ, respectively, $50$ and $90$ logarithmically spaced points in the spectral range $k_x \in [10^{-4},10^{0.48}]$ and $k_z \in [10^{-2},10^{1.2}]$ similar to those employed in \citet{Jovanovic2005}.

We compute $\|\mathcal{H}_{\nabla}\|_{\mu}$ in equation \eqref{eq:H_nabla_mu} for each wavenumber pair $(k_x,k_z)$ using the \texttt{mussv} command in the Robust Control Toolbox \citep{balas2005robust} of MATLAB R2020a. \rev{The arguments of \texttt{mussv} employed here include the state-space model of $\mathbfsbilow{H}_{\nabla}$ that samples the frequency domain adaptively\footnote{\rev{The command \texttt{mussv} can adaptively sample frequency domain $\omega\in \mathbb{R}^+$, and the frequency domain $\omega\in \mathbb{R}^-$ can be computed by modifying the state-space model}}. The \texttt{BlockStructure} argument comprises three full $N_y\times 3N_y$ complex matrices, and we use the \texttt{`Uf'} algorithm option.} The average computation time for each wavenumber pair $(k_x,k_z)$ is around $11s$ on a computer with a 3.4 GHz Intel Core i7-3770 CPU and 16GB RAM. These computations can be easily parallelized over either \rev{the} $k_x$ or $k_z$ domain  e.g., using the \texttt{parfor} command in the Parallel Computing Toolbox in MATLAB.

\vspace{2mm}

\section{Structured spatio-temporal frequency response}
\label{sec:result}

In this section, we use $\|\mathcal{H}_{\nabla}\|_{\mu}(k_x,k_z)$ in equation \eqref{eq:H_nabla_mu} to characterize the flow structures \rev{  (i.e., the ($k_x$,$k_z$) wavenumber pairs)} that are most amplified in transitional plane Couette flow and plane Poiseuille flow.  In order to illustrate the relative effect of the feedback interconnection \rev{versus} the imposed structure we compare the results to 
\begin{equation}
    \|\mathcal{H}\|_{\infty}(k_x,k_z):=\underset{\omega \in \mathbb{R}}{\text{sup}}\,\bar{\sigma}\left[\mathbfsbilow{H}(k_x,k_z,\omega)\right],\label{eq:H_inf}
    \end{equation}
 where $\mathbfsbilow{H}$ is the discretization of spatio-temporal frequency response operator $\mathcal{H}$ in \eqref{eq:linearized_ABC}. This quantity, which was previously analyzed   for transitional flows  \citep{jovanovic2004modeling,schmid2007nonmodal,illingworth2020streamwise}, describes the maximum singular value of the frequency response operator $\mathcal{H}$. \rev{This quantity} represents the maximal gain of $\mathcal{H}$ over all temporal frequencies, i.e., the worst-case amplification over harmonic inputs. Therefore the highest values of $\|\mathcal{H}\|_\infty$ correspond to structures that are most amplified but not those with the largest sustained energy density that is often reported in the literature, see e.g.,  \cite{Farrell1993,Bamieh2001,Jovanovic2005}.  
 
 In order to isolate the effect of the structure imposed on the nonlinearity from the effect of the closed-loop feedback interconnection, we also compute
\begin{equation}
    \|\mathcal{H}_{\nabla}\|_{\infty}(k_x,k_z):=\underset{\omega \in \mathbb{R}}{\text{sup}}\,\bar{\sigma}\left[\mathbfsbilow{H}_{\nabla}(k_x,k_z,\omega)\right].\label{eq:H_nabla_inf}
\end{equation}
This quantity is the unstructured counterpart of $\|\mathcal{H}_{\nabla}\|_{\mu}$, which is obtained by replacing the uncertainty set $\mathbfsbilow{\widehat{U}}_{\Upxi}$ with the set of full complex matrices $\mathbb{C}^{3N_y\times 9N_y}$ \citep{packard1993complex,zhou1996robust}. In other words, the definition does not specify a particular feedback pathway associated with each component of forcing, which leads to an unstructured feedback interconnection\footnote{ \rev{We note that by definition  $\|\mathcal{H}_{\nabla}\|_{\mu}(k_x,k_z)\leq \|\mathcal{H}_{\nabla}\|_{\infty}(k_x,k_z)$} \citep[equation (3.4)]{packard1993complex}.}. Comparisons between $\|\mathcal{H}_{\nabla}\|_{\mu}$ and $\|\mathcal{H}_{\nabla}\|_{\infty}$, therefore highlight the effect of \rev{ the structured uncertainty}. \rev{The values $\|\mathcal{H}\|_{\infty}$ in \eqref{eq:H_inf} and $\|\mathcal{H}_{\nabla}\|_\infty$ in \eqref{eq:H_nabla_inf} are computed using the \texttt{hinfnorm} command in the Robust Control Toolbox \citep{balas2005robust} of MATLAB. } 
 
\rev{ In the next subsection we analyze plane Couette flow at $Re=358$. This is followed by a study of plane Poiseuille flow at $Re=690$ in \S\ \ref{subsec:scale_dependent_mu_poi}. }\rev{These Reynolds numbers are within the ranges of $Re\in [340,393]$  and $Re\in [660,720]$, where oblique turbulent bands are respectively observed in plane Couette flow \citep{prigent2003long} and  plane Poiseuille flow \citep{kanazawa2018lifetime}. These  particular values were selected because there is data  from previous studies~\citep{prigent2003long,kanazawa2018lifetime} available for comparison. This section ends with a discussion of   the role of the componentwise structure of feedback interconnection} in the proposed structured input--output analysis in \S\ \ref{subsec:componentwise_structure_nonlinear}.

\subsection{Plane Couette flow at $Re=358$}
\label{subsec:scale_dependent_mu_cou}

\begin{figure}
	\hspace{0.05\textwidth}(a) \hspace{0.28\textwidth} (b) \hspace{0.28\textwidth} (c)

    \centering   
    
        \includegraphics[width=\textwidth]{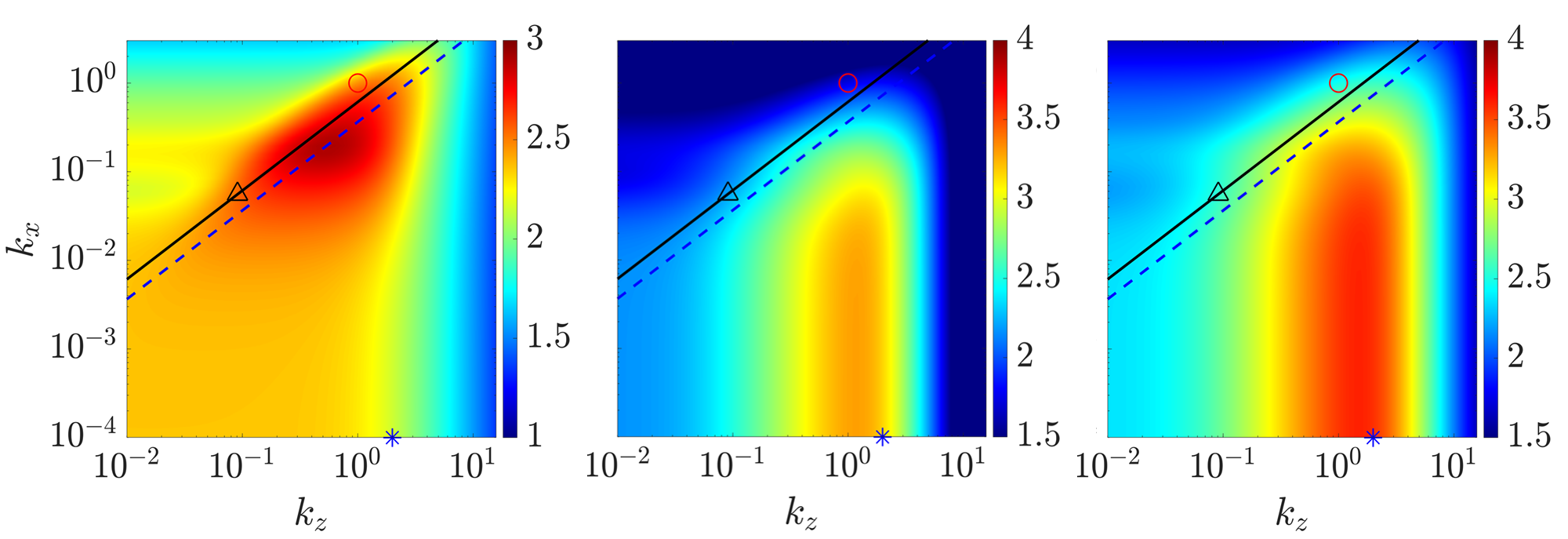}

    \caption{  (a) $\text{log}_{10}[\|\mathcal{H}_{\nabla}\|_{\mu}(k_x,k_z)]$, (b) $\text{log}_{10}[\|\mathcal{H}\|_{\infty}(k_x,k_z)]$, and (c)  $\text{log}_{10}[\|\mathcal{H}_{\nabla}\|_{\infty}(k_x,k_z)]$ for plane Couette flow at $Re=358$.  Here the symbols
    ({\color{blue}$\hspace{-0.0070in}\vspace{-0.015in}$\ding{83}}, blue) indicates streamwise vortices with $k_x\approx 0$, $k_z=2$;  ({\color{red}$\hspace{-0.0070in}\vspace{-0.015in}\Circle$}, red) marks oblique waves with $k_x=k_z=1$ \rev{as studied by \citep{reddy1998stability}. The symbol} ({\color{black}$\hspace{-0.0070in}\vspace{-0.015in}\triangle$}, black) marks $\lambda_x=113$, $\lambda_z=69$ which are the observed wavelengths of the oblique turbulent band at $Re=358$ \citep{prigent2003long}. The black solid line ($\mline\mline$) is $\lambda_z=\lambda_x\tan(32^\circ)$ representing a $32^\circ$ angle of the oblique turbulent band, \rev{and the blue dashed line ({\color{blue}$\dashed$}, blue) represents $\lambda_z=\lambda_x\tan(20^\circ)$.  }}
    \label{fig:mu_cou}
\end{figure}

In this subsection, we use the proposed approach to analyze the perturbations that are most likely to trigger transition in plane Couette flow at $Re=358$ using the $\|\mathcal{H}_{\nabla}\|_{\mu}$ formulation described in the previous section. Figure \ref{fig:mu_cou}(a) shows this quantity  alongside results obtained using an input--output analysis based approach describing the most amplified flow structures in terms of  $\|\mathcal{H}\|_{\infty}$ in panel (b)  and $\|\mathcal{H}_{\nabla}\|_{\infty}$ in equation \eqref{eq:H_nabla_inf} in panel (c). In all panels, we indicate the structures with $k_x\approx 0$ and $k_z=2$ representing streamwise vortices using ({\color{blue}$\hspace{-0.0070in}\vspace{-0.015in}$\ding{83}}, blue) and indicate $k_x=1$ and $k_z=1$ representing the oblique waves that were observed as the structures requiring the least energy to trigger transition in the DNS of \citet{reddy1998stability} using ({\color{red}$\hspace{-0.0070in}\vspace{-0.015in}\Circle$}, red). \rev{In general, the wavenumber pair $k_x\approx 0$ and $k_z=2$ marked by ({\color{blue}$\hspace{-0.0070in}\vspace{-0.015in}$\ding{83}}, blue) represents streamwise elongated flow structures that include both streamwise vortices and streamwise streaks. However, we will refer to ($k_x\approx 0,k_z=2$) as streamwise vortices when comparing with the results in  \citet{reddy1998stability} because that work explicitly introduced streamwise vortices associated with this wavenumber pair.} The figure shows clear differences in the dominant structures identified using the structured input--output approach. The largest magnitudes of $\|\mathcal{H}_{\nabla}\|_{\mu}$ in figure \ref{fig:mu_cou}(a) are associated with oblique waves with $k_x\in [10^{-2},1]$ and $k_z\in[10^{-1},1]$, while the streamwise \rev{elongated structures} that are dominant in panels (b) and (c) have a lesser but still large magnitude. This result is consistent with findings of \citet[figure 23]{reddy1998stability}  showing that oblique waves require less perturbation energy to trigger turbulence in plane Couette flow than streamwise vortices. A comparison of $\|\mathcal{H}_{\nabla}\|_{\infty}$ in figure \ref{fig:mu_cou}(c) and $\|\mathcal{H}\|_{\infty}$ in figure \ref{fig:mu_cou}(b) indicates that it is not the feedback interconnection that significantly changes the dominant flow structures but rather the imposition of the componentwise structure of the nonlinearity. In addition, we observe that the magnitude of $\|\mathcal{H}_{\nabla}\|_{\mu}$ in figure \ref{fig:mu_cou}(a) is lower than $\|\mathcal{H}_{\nabla}\|_{\infty}$ in figure \ref{fig:mu_cou}(c) for each $(k_x, k_z)$ pair, which is consistent with the fact that the unstructured gain $\|\mathcal{H}_{\nabla}\|_{\infty}$ provides an upper bound on the structured one, $\|\mathcal{H}_{\nabla}\|_{\mu}$ \citep{packard1993complex}.

The difference between the results in figure \ref{fig:mu_cou} mirrors the differences between the optimal perturbation structures predicted by linear and nonlinear optimal perturbation (NLOP) analysis. In particular, the structures predicted using $\|\mathcal{H}_\nabla\|_\mu$ are streamwise localized oblique waves reminiscent of those obtained as NLOP of plane Couette flow \citep{Monokrousos2011,duguet2010towards,Duguet2013,rabin2012triggering,cherubini2013nonlinear,cherubini2015minimal}, whereas the results obtained using $\|\mathcal{H}\|_\infty$ in figure \ref{fig:mu_cou}(b) indicate the dominance of the types of streamwise elongated flow structure\rev{s} predicted as linear optimal perturbations \citep{butler1992three}. Our results also reflect previous findings that the NLOP is wider in the spanwise direction than the linear optimal perturbation \rev{\citep[figure 11]{rabin2012triggering}}. The results in figure 
\ref{fig:mu_cou} therefore indicate that the current structured input--output framework provides closer agreement with both DNS and NLOP based predictions of perturbations to which the flow is most sensitive than traditional input--output methods focusing on the spatio-temporal frequency response $\mathcal{H}$. The inclusion of a feedback loop for $\|\mathcal{H}_{\nabla}\|_{\infty}$ in figure \ref{fig:mu_cou}(c) does lead to small improvements in the width of the structures predicted,  \rev{but it does not lead to identification of the dominance of the oblique waves. This behavior suggests that the \rev{weakening of the amplification of the streamwise elongated structures is a direct result of the structure imposed in the feedback interconnection}.}

Oblique turbulent bands have also been observed to be prominent in the transitional-regime of plane Couette flow with very large channel size \citep{prigent2003long,duguet2010formation}. Figure \ref{fig:mu_cou}  indicates the wavelength pair $\lambda_x=113$ and $\lambda_z=69$ ({\color{black}$\hspace{-0.0070in}\vspace{-0.015in}\triangle$}, black) associated with the  oblique turbulent bands \rev{that are observed to have horizontal extents in the range $\lambda_x\in[107,118]$ and $\lambda_z\in[62,76]$} in very large channel studies of plane Couette flow at $Re=358$, see \citep[figures 3(b) and 5]{prigent2003long}.
\rev{The characteristic inclination angle measured from the streamwise direction in $x-z$ plane is $\theta:=\tan^{-1}(\lambda_z/\lambda_x)=\tan^{-1}(69/113)\approx32^\circ$. This value is indicated by the black solid line $(\mline \mline)$: $\lambda_z=\lambda_x\tan(32^\circ)$ in figure \ref{fig:mu_cou} and falls within the mid-range of the angles $\theta\in [28^\circ,35^\circ]$ corresponding to the spread of the data in \citet[figure 5]{prigent2003long}. Other simulations employing a tilted domain to impose an angle constrained by periodic boundary conditions in the streamwise and spanwise directions indicate that the oblique turbulent bands can be maintained by an angle as low as $20^\circ$; see e.g., \citet[figure 6]{duguet2010formation}. In figure \ref{fig:mu_cou}, we also plot the angle $20^\circ$ represented by ({\color{blue} $\dashed$}, blue) $\lambda_z=\lambda_x\tan(20^\circ)$, which is shown to correspond to the center of the peak region of $\|\mathcal{H}_{\nabla}\|_\mu$. The results in the literature indicate that oblique structures associated with a range of wavelengths and inclination angles may provide large amplification, which may be the reason for the large peak region of $\|\mathcal{H}_{\nabla}\|_{\mu}$ in figure \ref{fig:mu_cou}(a). These results  suggest} that structured input--output analysis captures both \rev{the} wavelengths and \rev{the} angle of the oblique turbulent band in transitional plane Couette flow. While there is some footprint of these \rev{types of structures} in all three panels, the range of characteristic wavelengths and angles are most clearly associated with the peak region \rev{of}  $\|\mathcal{H}_{\nabla}\|_{\mu}$ in figure \ref{fig:mu_cou}(a), and the line representing the angle of the structures is quite consistent with the shape of the peak region. The fact that \rev{these structures become more prominent} through this analysis suggests that these turbulent bands arise in transitional flows due to their large amplification (sensitivity to disturbances).

\subsection{Plane Poiseuille flow at $Re=690$}

\label{subsec:scale_dependent_mu_poi}

\begin{figure}

	\hspace{0.05\textwidth}(a) \hspace{0.28\textwidth} (b) \hspace{0.28\textwidth} (c)

    \centering
    
    \includegraphics[width=\textwidth]{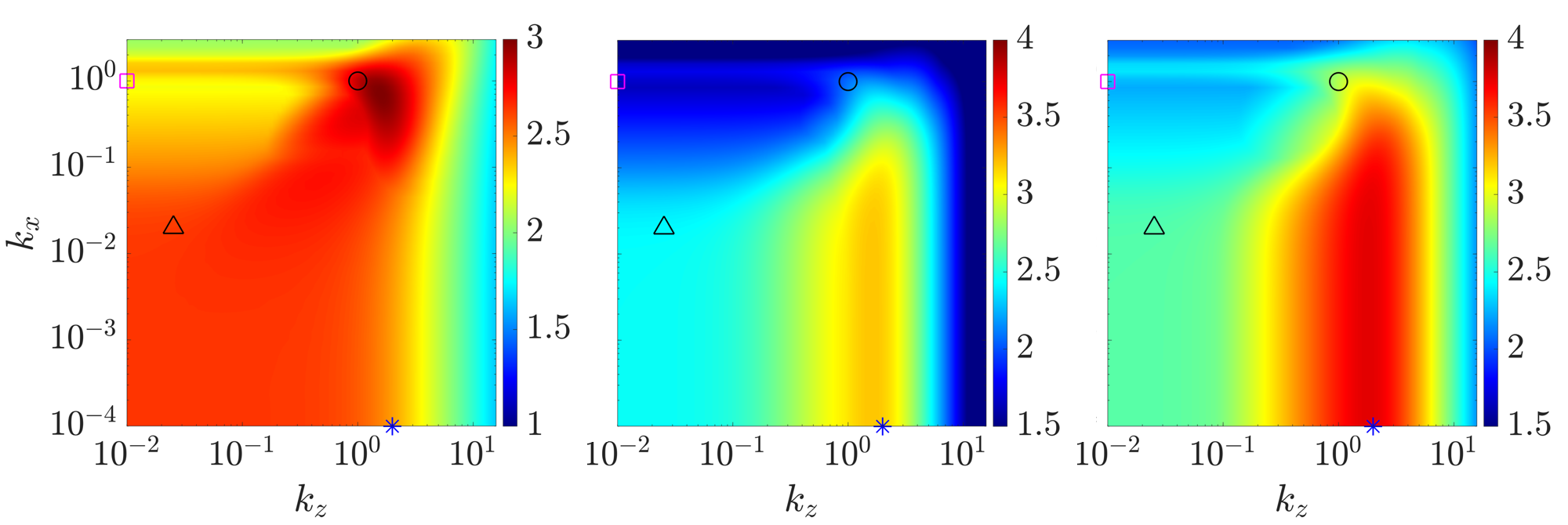}

    \caption{  (a) $\text{log}_{10}[\|\mathcal{H}_{\nabla}\|_{\mu}(k_x,k_z)]$, (b) $\text{log}_{10}[\|\mathcal{H}\|_{\infty}(k_x,k_z)]$, and (c) $\text{log}_{10}[\|\mathcal{H}_{\nabla}\|_{\infty}(k_x,k_z)]$ for plane Poiseuille flow at $Re=690$. Here the symbols
    ({\color{blue}$\hspace{-0.0070in}\vspace{-0.015in}$\ding{83}}, blue): $k_x\approx 0$, $k_z=2$ marks streamwise vortices; ({\color{black}$\hspace{-0.0070in}\vspace{-0.015in}\Circle$}, black): $k_x=k_z=1$ marks oblique waves;  ({\color{magenta}$\hspace{-0.0070in}\vspace{-0.015in}\square$}, magenta): $k_x=1$, $k_z\approx 0$ marks TS wave \rev{as studied by \citep{reddy1998stability}. The symbol} ({\color{black}$\hspace{-0.0070in}\vspace{-0.015in}\triangle$}, black): $\lambda_x=314$, $\lambda_z=248$ indicates the wavelengths of the oblique turbulent band at $Re=690$ observed in \citet{kanazawa2018lifetime}.  }
    \label{fig:mu_poi}
\end{figure}

In this subsection, we apply the proposed structured input--output analysis to investigate \rev{highly amplified} flow structures in plane Poiseuille flow at $Re=690$. Figure \ref{fig:mu_poi} compares (a)  $\|\mathcal{H_{\nabla}}\|_{\mu}$, (b)  $\|\mathcal{H}\|_{\infty}$, and (c) $\|\mathcal{H}_{\nabla}\|_\infty$ for this flow configuration. In each panel, we also indicate the streamwise vortices with $k_x\approx 0$ and $k_z=2$ ({\color{blue}$\hspace{-0.0070in}\vspace{-0.015in}$\ding{83}}, blue), oblique waves with $k_x=k_z=1$ ({\color{black}$\hspace{-0.0070in}\vspace{-0.015in}\Circle$}, black), and TS waves with $k_x=1$ and $k_z\approx 0$ ({\color{magenta}$\hspace{-0.0070in}\vspace{-0.015in}\square$}, magenta) that were identified as transition-inducing perturbations in \citet{reddy1998stability}. 
Similar to the results for plane Couette flow in figure \ref{fig:mu_cou},  the quantities $\|\mathcal{H}\|_{\infty}$ and $\|\mathcal{H}_\nabla\|_{\infty}$ show qualitatively similar behavior; the highest values for both correspond to streamwise streaks and vortices. \rev{In figures \ref{fig:mu_poi}(b) and \ref{fig:mu_poi}(c),} the TS wave structure appears as a local peak with a magnitude that is about an order of magnitude smaller than the values associated with the streamwise vortices in $\|\mathcal{H}_\nabla\|_\infty$ and $\|\mathcal{H}\|_\infty$. \rev{In these two panels,} the values for the oblique waves are of a similar order of magnitude as the peak corresponding to the TS waves in $\|\mathcal{H}\|_\infty$ and slightly higher in $\|\mathcal{H}_\nabla\|_\infty$. These findings agree with previous analyses of $\|\mathcal{H}\|_\infty$ in \citet{jovanovic2004modeling,schmid2007nonmodal}. \rev{The similarity of the $\|\mathcal{H}_\nabla\|_\infty$ and  $\|\mathcal{H}\|_\infty$} results indicate that an unstructured feedback interconnection  does not lead to substantial changes in the most prominent structures.

The overall shape of $\|\mathcal{H}_\nabla\|_\mu$ is somewhat different than that of either $\|\mathcal{H}\|_\infty$ or $\|\mathcal{H}_\nabla\|_\infty$. The streamwise \rev{elongated structures} that are dominant in panels (b) and (c) have a lesser but still large magnitude, while the peak value corresponds to oblique waves. The TS wave corresponds to a local peak in $\|\mathcal{H}_{\nabla}\|_{\mu}$, but the magnitudes are smaller than the peak values associated with oblique waves. This result is consistent with findings of \citet[figure 19]{reddy1998stability}  showing that oblique waves require slightly less perturbation energy to trigger turbulence in plane Poiseuille flow than streamwise vortices. Both the peak region and the large region of very high values in the bottom left quadrant of \rev{figure \ref{fig:mu_poi}(a)} are consistent with the short-timescale NLOP of plane Poiseuille flow, which was shown to be spatially localized with streamwise wavelength larger than spanwise wavelength \citep{farano2015hairpin}. These results indicate that the inclusion of structured uncertainty uncovers a broader range of transition-inducing structures and correctly orders their relative amplification in the sense of \rev{their} transition-inducing potential.

There is evidence that the oblique turbulent bands that are observed in very large channel studies also play a role in transition. Their ability to trigger transition has been exploited in a number of studies that employ flow  fields with  a sustained  oblique turbulent bands at a relatively high $Re$ as the initial conditions to trigger the banded turbulent-laminar patterns associated with transitioning flows at a Reynolds number of interest; see e.g., \citep{tsukahara2005dns,tuckerman2014turbulent,tao2018extended,xiao2020growth}. The characteristic wavelength pair ($\lambda_x=314$,  $\lambda_z=248$) associated with this structure  in plane Poiseuille flow at $Re=690$ (estimated from \citet[figure 5.1(b)]{kanazawa2018lifetime}) is indicated in each panel of figure \ref{fig:mu_poi} using ({\color{black}$\hspace{-0.0070in}\vspace{-0.015in}\triangle$}). These characteristic wavelengths are located within the range of large values of $\|\mathcal{H}_{\nabla}\|_\mu$. They are not associated with peak regions of  $\|\mathcal{H}\|_\infty$ or $\|\mathcal{H}_\nabla\|_\infty$ in figures \ref{fig:mu_poi}(b) or (c), although a footprint of these flow structures is visible in both. Figure \ref{fig:mu_poi}(a) indicates that the flow structure associated with the oblique turbulent band has a similar \rev{amplification under structured forcing} as streamwise elongated structures, although both of their magnitudes are smaller than \rev{that associated with the} oblique waves. Further analysis of these structures and their role in transition is a topic of ongoing work.

\subsection{Componentwise structure of nonlinearity: \rev{weakening} of the lift-up mechanism}

\label{subsec:componentwise_structure_nonlinear}

\begin{figure}
	\hspace{0.05\textwidth}(a) \hspace{0.28\textwidth} (b) \hspace{0.28\textwidth} (c)

\includegraphics[width=\textwidth]{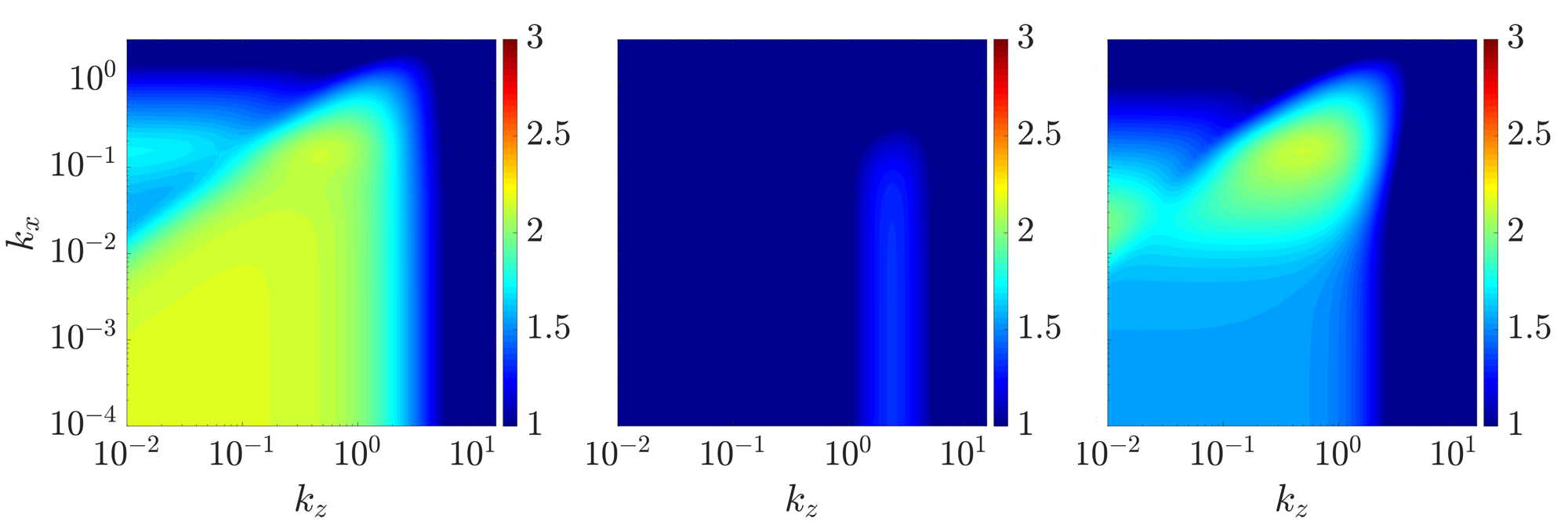}

	\hspace{0.05\textwidth}(d) \hspace{0.28\textwidth} (e) \hspace{0.28\textwidth} (f)

\includegraphics[width=\textwidth]{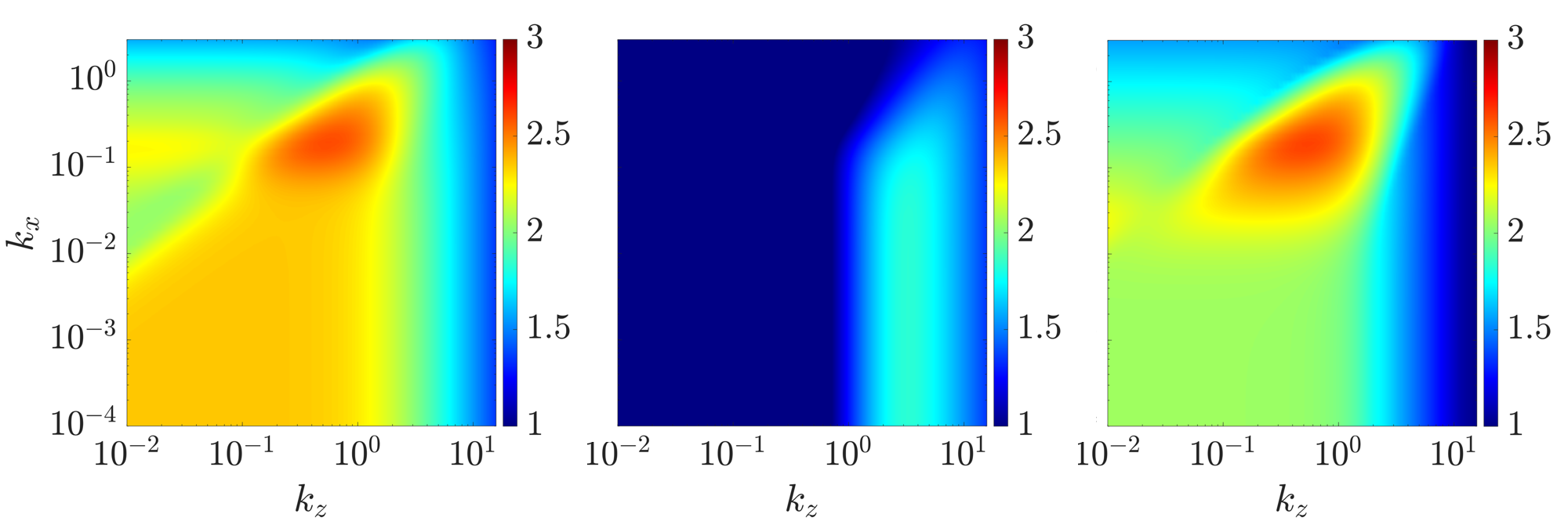}

    \caption{$\!$(a)$\,\text{log}_{10}[\|\mathcal{H}_{ux}\|_{\infty}(k_x,k_z)]$, (b)$\,\text{log}_{10}[\|\mathcal{H}_{vy}\|_{\infty}(k_x,k_z)]$, (c)$\,\text{log}_{10}[\|\mathcal{H}_{wz}\|_{\infty}(k_x,k_z)]$, (d)$\,\text{log}_{10}[\|\mathcal{H}_{\nabla {ux}}\|_{\infty}(k_x,k_z)]$, (e)$\,\text{log}_{10}[\|\mathcal{H}_{\nabla {vy}}\|_{\infty}(k_x,k_z)]$, and (f)$\,\text{log}_{10}[\|\mathcal{H}_{\nabla {wz}}\|_{\infty}(k_x,k_z)]$ for plane Couette flow at $Re=358$.}
    \label{fig:H_inf_no_grad_xu_yv_zw_cou}
\end{figure}

\begin{figure}
	\hspace{0.05\textwidth}(a) \hspace{0.28\textwidth} (b) \hspace{0.28\textwidth} (c)

\includegraphics[width=\textwidth]{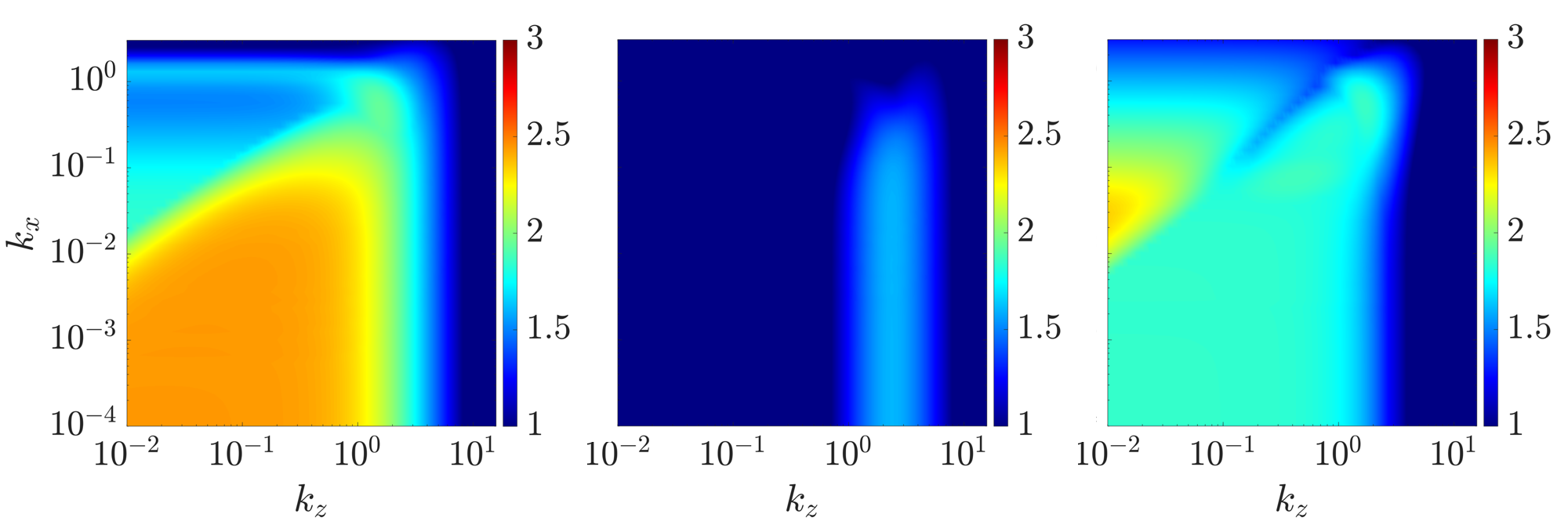}

	\hspace{0.05\textwidth}(d) \hspace{0.28\textwidth} (e) \hspace{0.28\textwidth} (f)

\includegraphics[width=\textwidth]{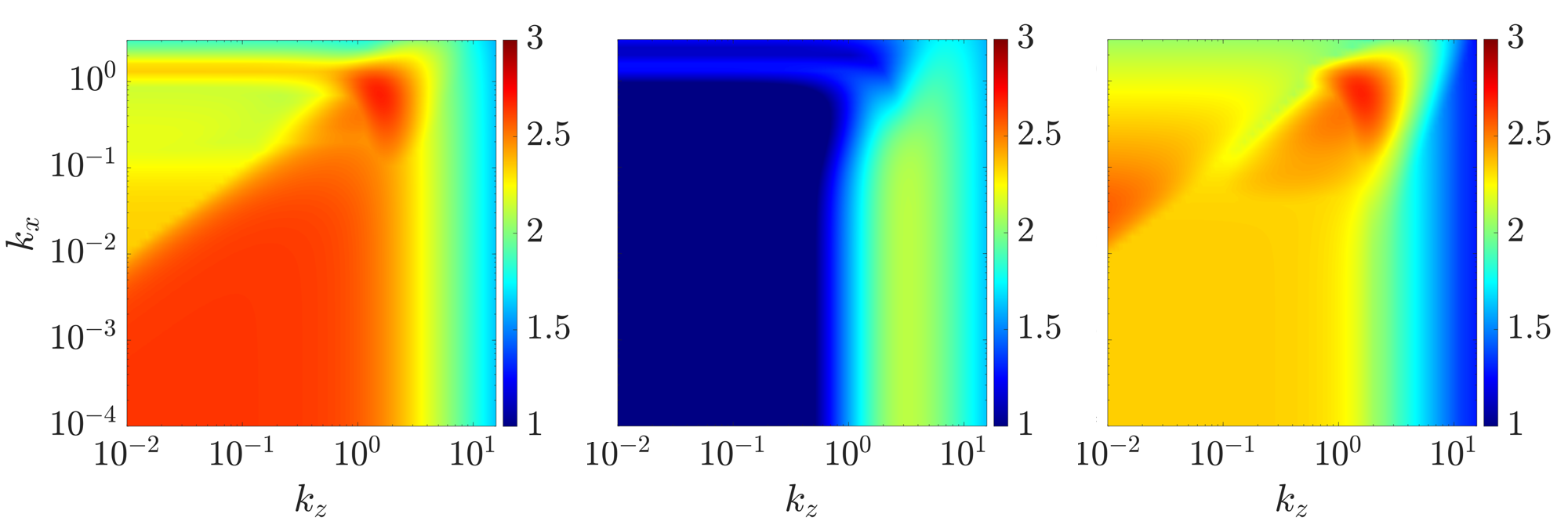}

    \caption{$\!$(a)$\,\text{log}_{10}[\|\mathcal{H}_{ux}\|_{\infty}(k_x,k_z)]$, (b)$\,\text{log}_{10}[\|\mathcal{H}_{vy}\|_{\infty}(k_x,k_z)]$, (c)$\,\text{log}_{10}[\|\mathcal{H}_{wz}\|_{\infty}(k_x,k_z)]$, (d)$\,\text{log}_{10}[\|\mathcal{H}_{\nabla {ux}}\|_{\infty}(k_x,k_z)]$, (e)$\,\text{log}_{10}[\|\mathcal{H}_{\nabla {vy}}\|_{\infty}(k_x,k_z)]$, and (f)\,$\text{log}_{10}[\|\mathcal{H}_{\nabla {wz}}\|_{\infty}(k_x,k_z)]$ for plane Poiseuille flow at $Re=690$.}
    \label{fig:H_inf_no_grad_xu_yv_zw_poi}
\end{figure}

The results in the previous subsections, particularly the differences between $\|\mathcal{H}_\nabla\|_\mu$ and  $\|\mathcal{H}_\nabla\|_\infty$ in figures \ref{fig:mu_cou} and \ref{fig:mu_poi} highlight the role of the feedback interconnection structure in the identification of the perturbations to which the flow is most sensitive. In particular, the imposition of the componentwise structure leads to lesser prominence of streamwise elongated structures $(k_x\approx 0, k_z=2)$ in  $\|\mathcal{H}_{\nabla}\|_{\mu}$ versus 
$\|\mathcal{H}\|_{\infty}$  or $\|\mathcal{H}_{\nabla}\|_{\infty}$ in both plane Couette and Poiseuille flows (see figures \ref{fig:mu_cou} and \ref{fig:mu_poi}).

The mechanisms underlying the differences in $\|\mathcal{H}_{\nabla}\|_{\mu}$ and $\|\mathcal{H}_{\nabla}\|_{\infty}$
can be analyzed by isolating the effect of forcing in each component of the momentum equation, i.e. $f_x,\, f_y, \, f_z$ in equation \eqref{eq:f_nonlinear} on the amplification of each velocity component $u$, $v$, $w$. These nine quantities are associated with $\|\mathcal{H}_{ij}\|_{\infty}$, where the spatio-temporal frequency response operator $\mathcal{H}_{ij}$ from each forcing component ($j=x,y,z$) to each velocity component ($i=u,v,w$) is given by \citep{Jovanovic2005}
\begin{equation}\mathcal{H}_{ij}=\widehat{\mathcal{C}}_{i}\left(\text{i}\omega \mathcal{I}_{2\times 2}-\widehat{\mathcal{A}}\right)^{-1}\widehat{\mathcal{B}}_{j}\end{equation}
with 
\begingroup
\allowdisplaybreaks
\begin{subequations}
\label{eq:H_componentwise}
\begin{align}
     \mathcal{\widehat{B}}_x:=&\mathcal{\widehat{B}}\begin{bmatrix}
    \mathcal{I} & 0 & 0
    \end{bmatrix}^{\text{T}},   \;\; \mathcal{\widehat{B}}_y:=\mathcal{\widehat{B}}\begin{bmatrix}
    0 & \mathcal{I} & 0
    \end{bmatrix}^{\text{T}},\;\;\mathcal{\widehat{B}}_z:=\mathcal{\widehat{B}}\begin{bmatrix}
    0 & 0 & \mathcal{I}
    \end{bmatrix}^{\text{T}},\\ \mathcal{\widehat{C}}_u:=&\begin{bmatrix}\mathcal{I} & 0 & 0
    \end{bmatrix}   \mathcal{\widehat{C}} ,\;\;\;\;\;\mathcal{\widehat{C}}_v:=\begin{bmatrix}0 & \mathcal{I} & 0
    \end{bmatrix}   \mathcal{\widehat{C}},
    \;\;\;\;\;\mathcal{\widehat{C}}_w:=\begin{bmatrix}0 & 0 & \mathcal{I}
    \end{bmatrix}   \mathcal{\widehat{C}}.
\end{align}
\end{subequations}
\endgroup

These quantities were analyzed in \citep{jovanovic2004modeling} and \citep{schmid2007nonmodal}. Those results indicate that the most significant amplification is seen when forcing is applied in the cross-stream and the output is the streamwise velocity component; i.e., that associated with respective frequency response operators $\mathcal{H}_{uy}$, $\mathcal{H}_{uz}$ and input--output \rev{pathway} $f_y\rightarrow u$, $f_z\rightarrow u$. Similar behavior occurs if we isolate $\mathcal{H}_{\nabla}$ by examining each input--output response pathways:
\begin{align}
    \mathcal{H}_{\nabla ij}:=\boldsymbol{\widehat{\nabla}}\mathcal{H}_{ij}. 
    \label{eq:H_nabla_ij_def}
\end{align}
In this prior work, the input forcing (applied either directly to the LNS (top-block in figure \ref{fig:feedback_introduction}) or through a feedback interconnection) was unstructured in the sense that there was no restriction in terms of the permissible input--output pathways. \rev{The} behavior of the largest $\|\mathcal{H}_{ij}\|_\infty$ ($\|\mathcal{H}_{\nabla {ij}}\|_\infty$) response \rev{therefore} dominates the overall $\|\mathcal{H}\|_\infty$ ($\|\mathcal{H}_\nabla\|_\infty$) response. 

The structured input--output analysis framework introduced here instead imposes a correlation between each component of the modeled forcing $f_{x,\xi}$, $f_{y,\xi}$, and $f_{z,\xi}$ and the respective velocity components $u$, $v$, and $w$ by constraining the feedback interconnection to retain the componentwise structure of our input--output model of the forcing. This model of the forcing in terms of componentwise input--output relationships from $\boldsymbol{\nabla} u$, $\boldsymbol{\nabla} v$, $\boldsymbol{\nabla} w$ to the respective components $f_{x,\xi}=-\boldsymbol{u}_{\xi}\cdot \boldsymbol{\nabla} u$, $f_{y,\xi}=-\boldsymbol{u}_{\xi}\cdot \boldsymbol{\nabla} v$, and $f_{z,\xi}=-\boldsymbol{u}_{\xi}\cdot \boldsymbol{\nabla} w$ with the gain defined in terms of $-\boldsymbol{u}_{\xi}$ constrains the feedback relationships such that  each component of the forcing 
is most strongly influenced by that component of the velocity field and velocity gradient. 
 These constraints on the permissible feedback pathways within our model of the nonlinear interactions limit the influence of the input--output pathways $f_y\rightarrow u$ and $f_z\rightarrow u$. \rev{The structured input--output response $\|\mathcal{H}_{\nabla}\|_\mu$ is instead associated with input--output pathways} $f_x\rightarrow u$, $f_y\rightarrow v$, and $f_z\rightarrow w$ \rev{as} illustrated in figures \ref{fig:H_inf_no_grad_xu_yv_zw_cou} and  \ref{fig:H_inf_no_grad_xu_yv_zw_poi}, which respectively plot (a) $\|\mathcal{H}_{ux}\|_{\infty}$, (b) $\|\mathcal{H}_{vy}\|_{\infty}$, (c) $\|\mathcal{H}_{wz}\|_{\infty}$, (d) $\|\mathcal{H}_{\nabla ux}\|_{\infty}$, (e) $\|\mathcal{H}_{ \nabla vy}\|_{\infty}$, (f) $\|\mathcal{H}_{\nabla wz}\|_{\infty}$  for the plane Couette and Poiseuille cases respectively discussed in \S\ \ref{subsec:scale_dependent_mu_cou} and \S\ \ref{subsec:scale_dependent_mu_poi}. Here, we can see that the results of structured input--output analysis $\|\mathcal{H}_{\nabla}\|_{\mu}$ for both of these flows in figures \ref{fig:mu_cou}(a) and \ref{fig:mu_poi}(a) resemble the combined effect of this limited set of input--output pathways. \rev{Moreover, the quantity $\|\mathcal{H}_{\nabla}\|_{\mu}$ at each wavenumber pair $(k_x,k_z)$ is lower bounded by $\|\mathcal{H}_{\nabla ux}\|_{\infty}$, $\|\mathcal{H}_{\nabla vy}\|_{\infty}$, and $\|\mathcal{H}_{\nabla wz}\|_{\infty}$ as described in theorem \ref{lemma:mu_componentwise_inf}, whose proof is provided in Appendix \ref{appendix:mu_componentwise_inf}. The relationship in theorem  \ref{lemma:mu_componentwise_inf} is evident when comparing results in figures \ref{fig:mu_cou}(a) and \ref{fig:H_inf_no_grad_xu_yv_zw_cou}(d)-(f) for plane Couette flow and comparing results in figures \ref{fig:mu_poi}(a) and \ref{fig:H_inf_no_grad_xu_yv_zw_poi}(d)-(f) for plane Poiseuille flow. }

 \rev{\begin{thm}
Given wavenumber pair $(k_x,k_z)$. 
\begin{align}
    \|\mathcal{H}_{\nabla}\|_{\mu}\geq \text{max}[\|\mathcal{H}_{\nabla ux}\|_{\infty},\|\mathcal{H}_{\nabla vy}\|_{\infty}, \|\mathcal{H}_{\nabla wz}\|_{\infty}].
    \label{eq:mu_larger_than_all_diagonal_component}
\end{align}
\label{lemma:mu_componentwise_inf}
\end{thm} }

 The input--output pathways that dominate the overall unstructured response $\|\mathcal{H}\|_{\infty}$ ($\|\mathcal{H}_{uy}\|_{\infty}$ and $\|\mathcal{H}_{uz}\|_{\infty}$) emphasize amplification of streamwise streaks by cross-stream forcing; i.e., the lift-up mechanism, see e.g., the discussion in  \citet{jovanovic2020bypass} for further details.  The lift-up mechanism therefore appears to be weakened through the imposition of the componentwise structure of the nonlinearity, which is consistent with results suggesting that nonlinear mechanisms \rev{disadvantage the growth of streaks}, see e.g, \citet{Duguet2013,Brandt2014}. These results suggest that the preservation of the componentwise structure of nonlinearity within the proposed approach enables the method to capture important nonlinear effects, leading to better agreement with DNS and experimental studies and nonlinear analysis of the perturbations that require less energy to initiate transition, e.g. NLOP.

\rev{
\section{Reynolds number dependence}
\label{sec:Re_dependence}
}

In this section, we aggregate results across a range of $(k_x,k_z)$ scales to study the Reynolds number dependence and the associated scaling law of $\|\mathcal{H}_{\nabla}\|_{\mu}$ for both plane Couette flow and plane Poiseuille flows. In particular, we compute
\begin{align}
    \|\mathcal{H}_{\nabla}\|_{\mu}^M:=&\underset{k_z,\,k_x}{\text{max}}\|\mathcal{H}_{\nabla}\|_{\mu}(k_x,k_z), 
    \label{eq:H_nabla_mu_M}
\end{align}
where $\underset{k_z,\,k_x}{\text{max}}$ corresponds to the maximum value over the  wavenumber pair\rev{s} $(k_x,k_z)$ in the computational range of $k_x \in [10^{-4},10^{0.48}]$ and $k_z \in [10^{-2},10^{1.2}]$. 

In order to compare our results to the scaling relationships of  $\|\mathcal{H}\|_{\infty}$ previously described in the literature and to isolate the effect of the structure in the feedback loop, we analogously define
\begin{subequations}
\label{eq:H_inf_and_H_nabla_inf_M}
\begin{align}
    \|\mathcal{H}\|_{\infty}^M:=&\underset{\substack{k_z,\,k_x}}{\text{max}}\,\|\mathcal{H}\|_{\infty}(k_x,k_z),    \label{eq:H_inf_M}\\
    \|\mathcal{H}_{\nabla}\|_{\infty}^M:=&\underset{\substack{k_z,\,k_x}}{\text{max}}\,\|\mathcal{H}_{\nabla}\|_{\infty}(k_x,k_z). 
    \label{eq:H_nabla_inf_M}
\end{align} 
\end{subequations}
The scaling of quantities related to $\|\mathcal{H}\|_{\infty}$ and $ \|\mathcal{H}\|_{\infty}^M$ with different input--output pathways, i.e. different $\widehat{\mathcal{B}}$ and $\widehat{\mathcal{C}}$ matrices in equation \eqref{eq:H_componentwise}, has been widely studied. For example, \citet[table 1]{trefethen1993hydrodynamic} showed that $\underset{\omega\in \mathbb{R}}{\text{sup}}\|(\text{i}\omega\mathcal{I}-\widehat{\mathcal{A}})^{-1}\|\sim Re^2$ \rev{for plane Couette flow and plane  Poiseuille flow are respectively associated with wavenumber pairs $(k_x,k_z)=(0,1.18)$ and $(k_x,k_z)=(0,1.62)$.} Here, the operator norm $\|\cdot\|$ \rev{is defined such that} $\underset{\omega\in \mathbb{R}}{\text{sup}}\|\cdot\|$ is equivalent to the definition of $\|\cdot\|_{\infty}$ employed in \eqref{eq:H_inf}. \citet{kreiss1994bounds} showed that the related quantity \rev{maximized over a range of $(k_x,k_z)$, i.e.,} \[ \underset{\mathbb{R}e[s]\geq0}{\text{max}}\|(s\mathcal{I}-\widehat{\mathcal{A}})^{-1}\|\sim Re^2\] for plane Couette flow, where $\mathbb{R}e[s]$ denotes the real part of Laplace variable $s$. \citet[theorem 11]{jovanovic2004modeling} analytically derived the same $\sim Re^2$ scaling for the special case of $\|\mathcal{H}\|_\infty$ restricted to $k_x=0$ for both plane Couette and Poiseuille flows.

\begin{figure}

	(a) \hspace{2.42in} (b) 

    \centering
    \includegraphics[width=2.6in]{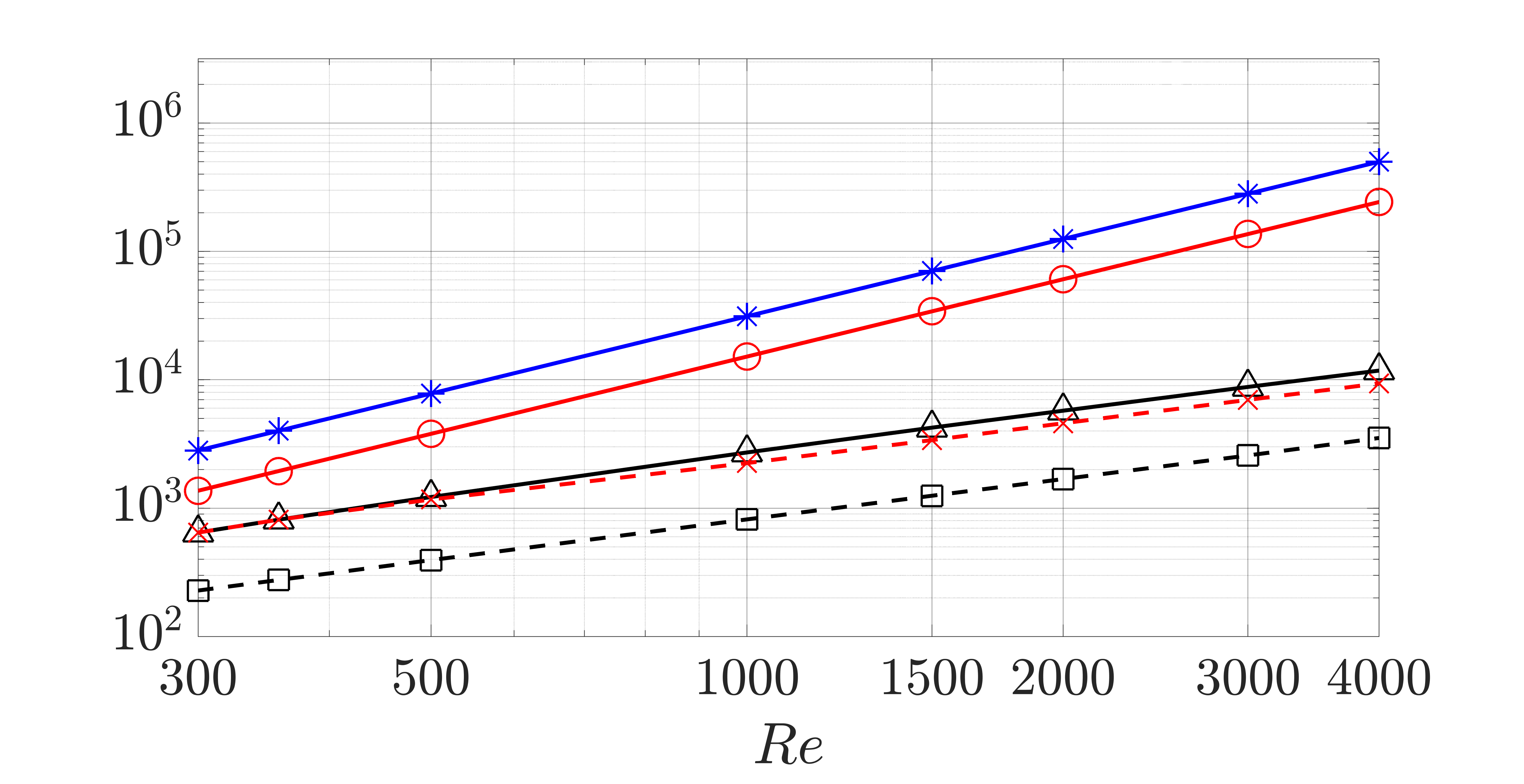}
    \includegraphics[width=2.6in]{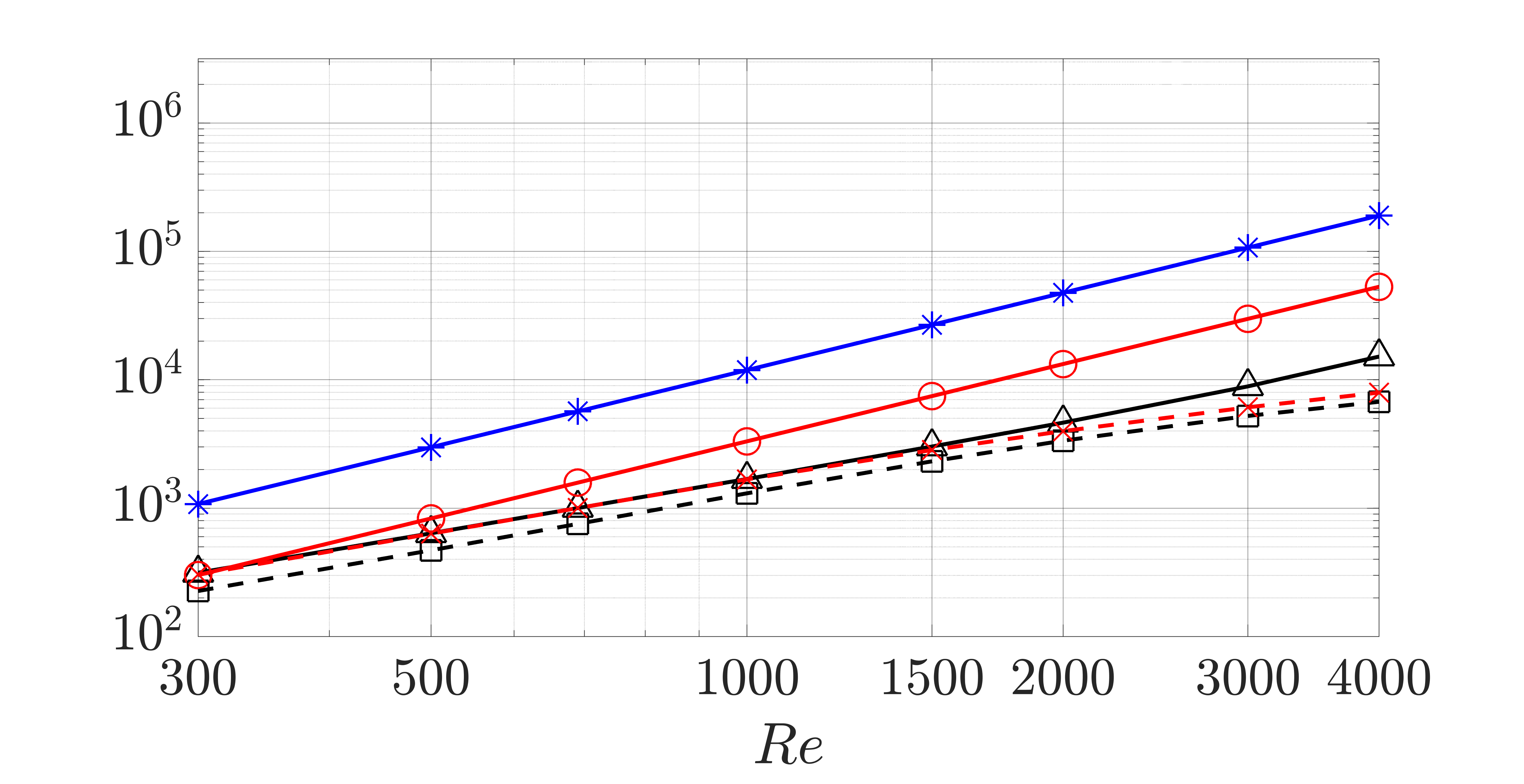}

    \caption{\setstretch{1.2}The Reynolds number dependence of $\|\mathcal{H}_{\nabla}\|_{\mu}^M$ ({\color{black}$\hspace{-0.0070in}\vspace{-0.015in}\triangle\mline\mline\mline$}, black); $\|\mathcal{H}\|_{\infty}^M$ ({\color{red}$\hspace{-0.0070in}\vspace{-0.015in}\Circle\mline\mline\mline$}, red); $\|\mathcal{H}_{\nabla}\|_{\infty}^M$ ({\color{blue}$\hspace{-0.0070in}\vspace{-0.015in}$\ding{83}$\mline\mline\mline$}, blue); $\|\mathcal{H}_{\nabla}\|_{\mu}(1,1)$ ({\color{black}$\hspace{-0.0070in}\vspace{-0.015in}\square\dashed$}, black). Here, panel (a) is plane Couette flow with ({\color{red}$\hspace{-0.0070in}\vspace{-0.015in}\times\dashed$}, red) marks $\|\mathcal{H}_{\nabla}\|_{\mu}(0.19,0.58)$ and panel (b) is plane Poiseuille flow with ({\color{red}$\hspace{-0.0070in}\vspace{-0.015in}\times\dashed$}, red) marks $\|\mathcal{H}_{\nabla}\|_{\mu}(0.69,1.56)$. }
    \label{fig:Re_dependence}
\end{figure}

\begin{figure}

	(a) \hspace{2.42in} (b) 

    \centering
    
    \includegraphics[width=2.6in]{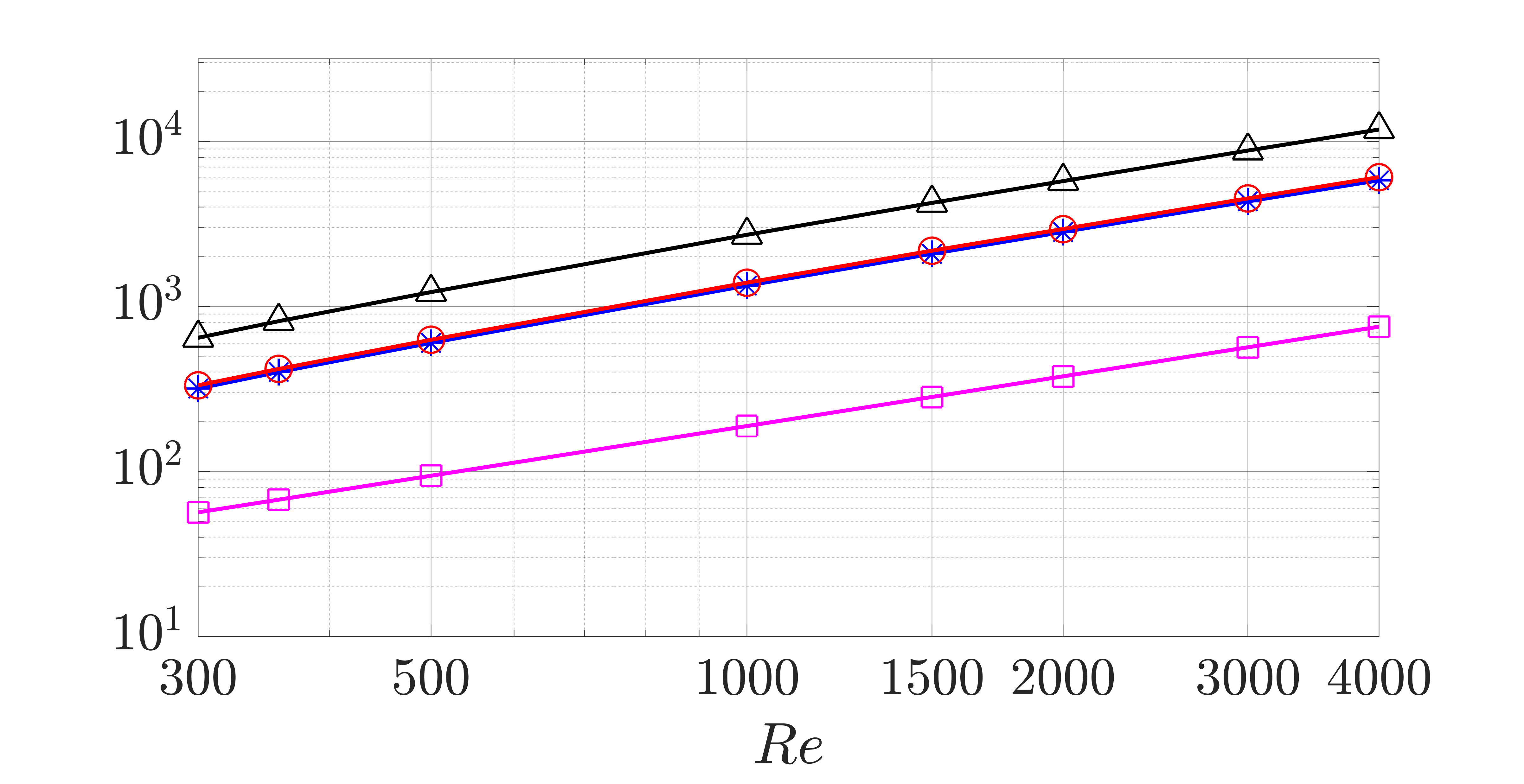}
     \includegraphics[width=2.6in]{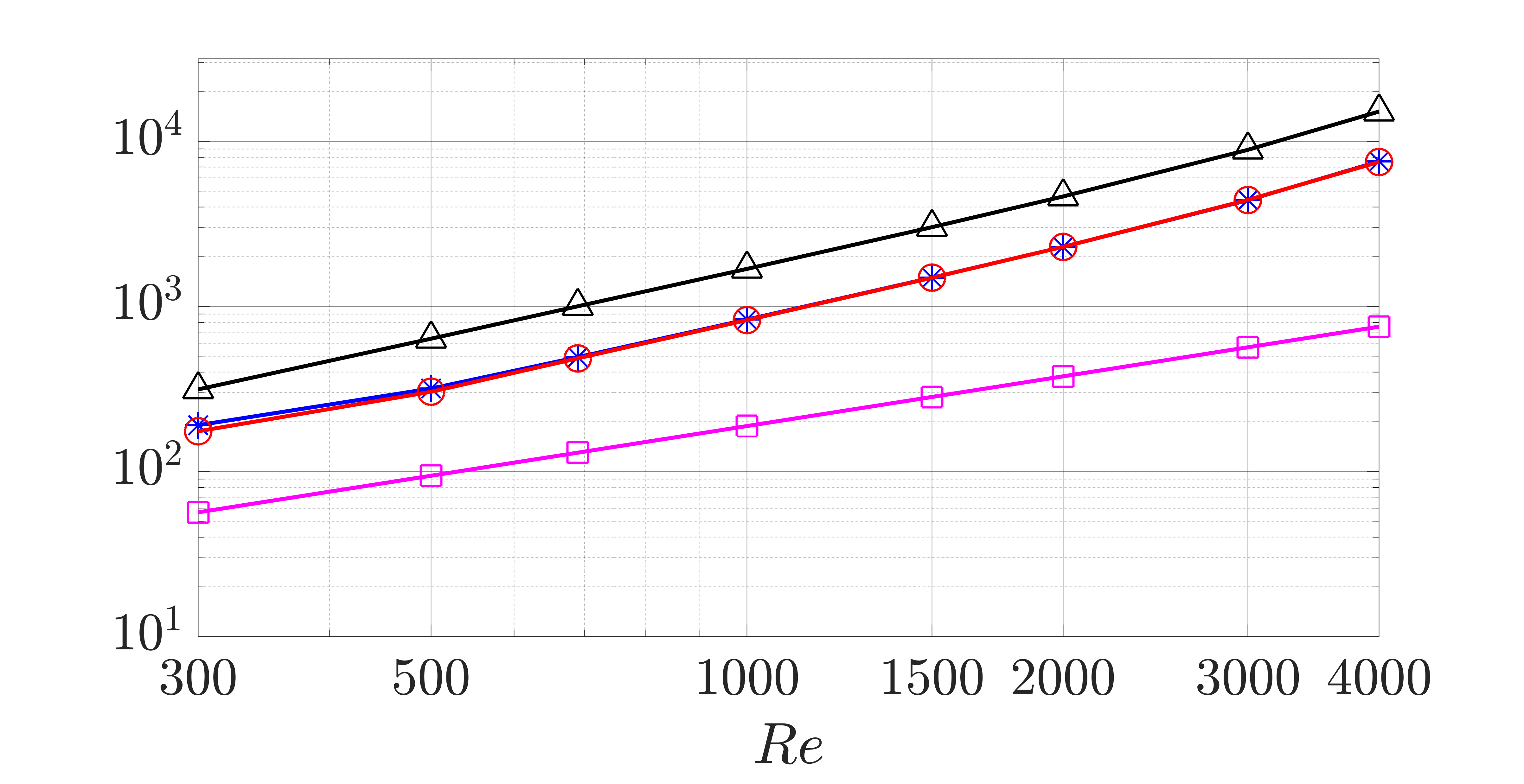}
    \caption{\setstretch{1.25}The Reynolds number dependence of $\|\mathcal{H}_{\nabla}\|_{\mu}^M$ ({\color{black}$\hspace{-0.0070in}\vspace{-0.015in}\triangle\mline\mline\mline$}, black); $\|\mathcal{H}_{\nabla ux}\|_{\infty}^M$ ({\color{blue}$\hspace{-0.0070in}\vspace{-0.015in}$\ding{83}$\mline\mline\mline$}, blue); $\|\mathcal{H}_{\nabla vy}\|_{\infty}^M$  ({\color{magenta}$\hspace{-0.0070in}\vspace{-0.015in}\square\mline\mline\mline$}, magenta);
 $\|\mathcal{H}_{\nabla wz}\|_{\infty}^M$ ({\color{red}$\hspace{-0.0070in}\vspace{-0.015in}\Circle\mline\mline\mline$}, red). Here, panel (a) is plane Couette flow and panel (b) is plane Poiseuille flow.}
    \label{fig:Re_dependence_isolated}
\end{figure}

\begin{figure}

	(a) \hspace{0.49\textwidth} (b) 

    \centering
    
    \includegraphics[width=0.49\textwidth]{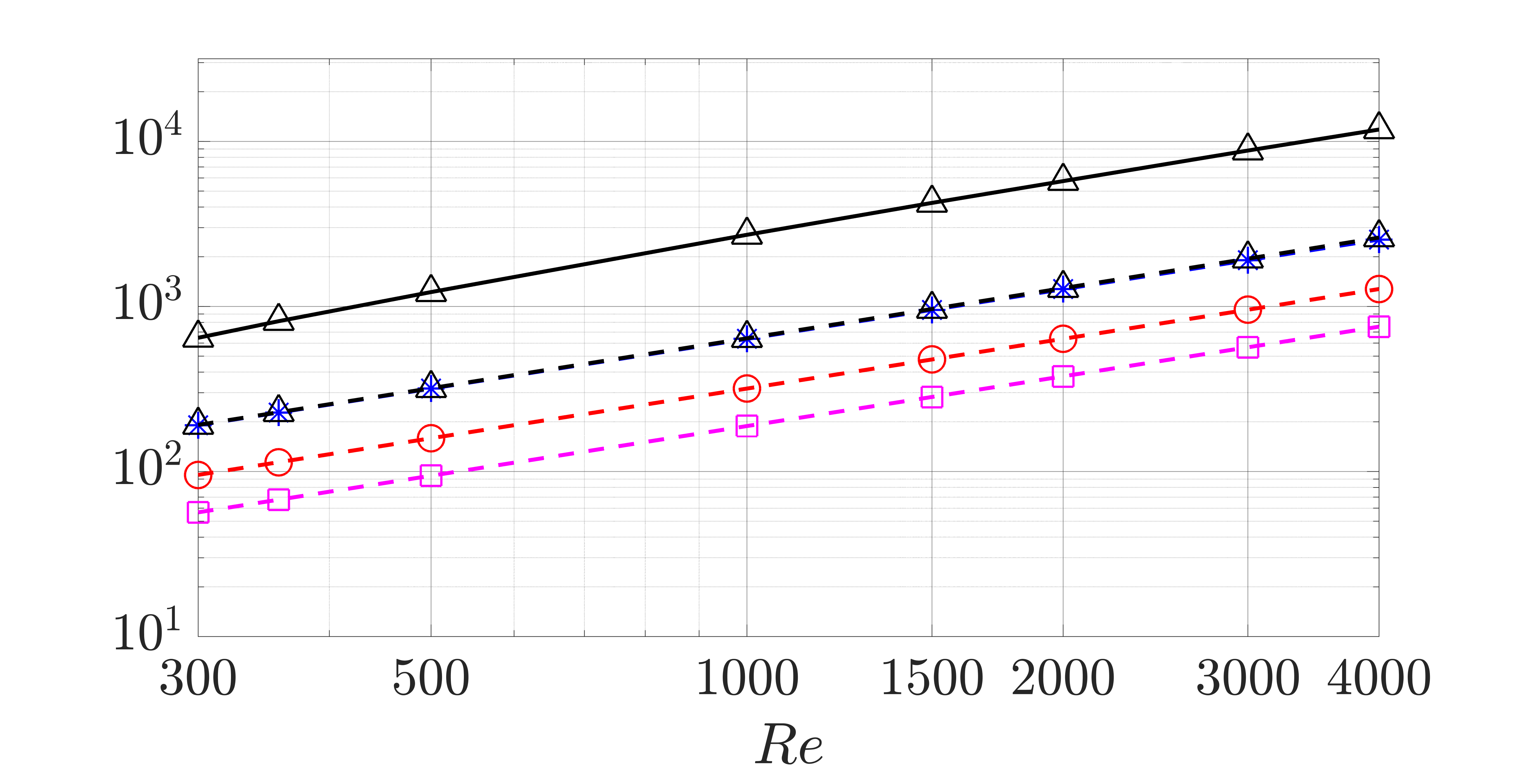}
     \includegraphics[width=0.49\textwidth]{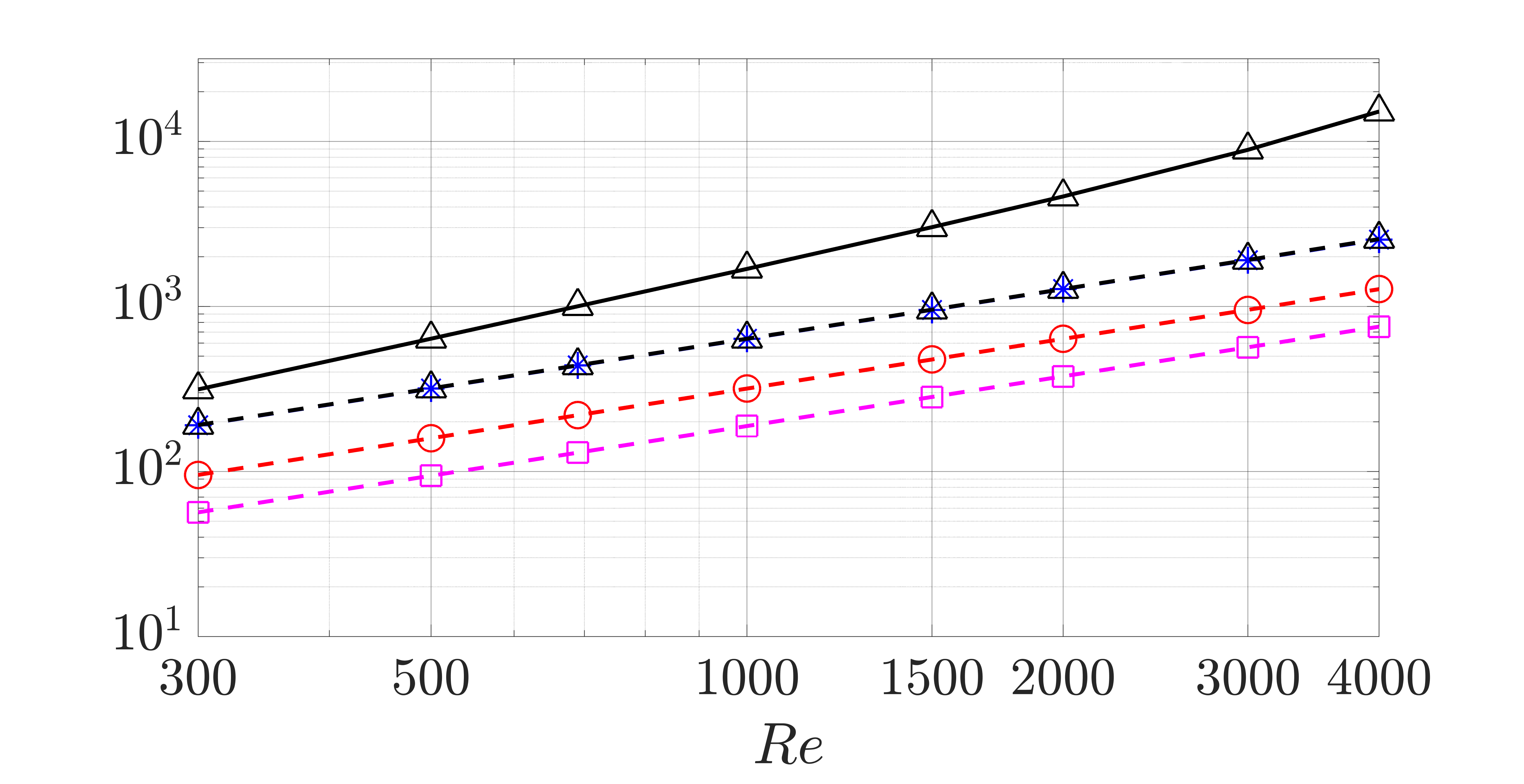}
    \caption{\setstretch{1.25}The Reynolds number dependence of $\|\mathcal{H}_{\nabla}\|_{\mu}^M$ ({\color{black}$\hspace{-0.0070in}\vspace{-0.015in}\triangle\mline\mline\mline$}, black); $\|\mathcal{H}_{\nabla}\|_{\mu}^{sc}$ ({\color{black}$\hspace{-0.0070in}\vspace{-0.015in}\triangle\dashed$}, black); $\|\mathcal{H}_{\nabla ux}\|_{\infty}^{sc}$  ({\color{blue}$\hspace{-0.0070in}\vspace{-0.015in}$\ding{83}$\dashed$}, blue); $\|\mathcal{H}_{\nabla vy}\|_{\infty}^{sc}$  ({\color{magenta}$\hspace{-0.0070in}\vspace{-0.015in}\square\dashed$}, magenta); $\|\mathcal{H}_{\nabla wz}\|_{\infty}^{sc}$ ({\color{red}$\hspace{-0.0070in}\vspace{-0.015in}\Circle\dashed$}, red). Here, panel (a) is plane Couette flow and panel (b) is plane Poiseuille flow.}
    \label{fig:Re_dependence_isolated_kx0}
\end{figure}

Figure \ref{fig:Re_dependence} plots the quantities in equations \eqref{eq:H_nabla_mu_M}-\eqref{eq:H_inf_and_H_nabla_inf_M} as a function of Reynolds number $(Re\in[300,4000])$ for (a) plane Couette flow and (b) plane Poiseuille flow. The upper bound of $Re=4000$ was selected to remain below the known  linear stability limit for plane Poiseuille flow of $Re\simeq 5772$ \citep{orszag1971accurate}. As expected all of these quantities increase with the Reynolds number and the values of $\|\mathcal{H}_\nabla\|_{\infty}^M$ are larger than those of $\|\mathcal{H}_{\nabla}\|_{\mu}^M$. We obtain a Reynolds number scaling of each quantity by fitting the lines in figure \ref{fig:Re_dependence} to $ c_0Re^{\eta}$, where $c_0$ is a constant scalar \rev{and $\eta$ is the corresponding scaling exponent}. The results show that  $\|\mathcal{H}\|_{\infty}^M$ and $\|\mathcal{H}_{\nabla}\|_{\infty}^M$ scale as $\sim Re^2$ in the range $Re\in[300,4000]$ for both plane Couette and plane Poiseuille flows. This scaling is consistent with the results in \citet{trefethen1993hydrodynamic} for the frequency response operator with identity operators for $\widehat{\mathcal{C}}$ and $\widehat{\mathcal{B}}$ as well as the related quantity in \citet{kreiss1994bounds}. The fact that the scaling of this quantity for the modified frequency response operator $\mathcal{H}_{\nabla}$ is the same as that of $\mathcal{H}$ suggests that adding an unstructured uncertainty in the feedback loop to represent the nonlinear interactions does not change the Reynolds number scaling.

The quantity $\|\mathcal{H}_{\nabla}\|_{\mu}^M$ in figure \ref{fig:Re_dependence} instead shows scalings of $\sim Re^{1.1}$ over $Re\in[300,4000]$ for plane Couette flow and $\sim Re^{1.5}$ for plane Poiseuille flow in the range $Re\in [500,4000]$. The difference between the scaling of $\|\mathcal{H}_{\nabla}\|_{\mu}^M$ and the $\sim Re^2$ scaling associated with either $\|\mathcal{H}\|_{\infty}^M$ or $\|\mathcal{H}_{\nabla}\|_{\infty}^M$ again arises through the imposition of the componentwise structure of nonlinearity. As discussed in \S\ \ref{subsec:componentwise_structure_nonlinear},  the reduced scaling is related to the smaller amplification in the input--output pathways $f_x\rightarrow u$, $f_y\rightarrow v$, and $f_z\rightarrow w$ and associated $f_x\rightarrow \boldsymbol{\nabla} u$, $f_y\rightarrow \boldsymbol{\nabla} v$, and $f_z\rightarrow \boldsymbol{\nabla} w$. The scaling for these input--output pathways can be evaluated directly through 
\begin{align}
    \|\mathcal{H}_{\nabla ij}\|_{\infty}^{M}:=&\underset{k_z,\,k_x}{\text{max}}\|\mathcal{H}_{\nabla ij}\|_{\infty}(k_x,k_z). 
\end{align}
\rev{These quantities are} plotted in figure \ref{fig:Re_dependence_isolated} alongside  $\|\mathcal{H}_{\nabla}\|_{\mu}^M$.  Performing a similar fit we find that both $\|\mathcal{H}_{\nabla ux}\|_{\infty}^M$ and $\|\mathcal{H}_{\nabla wz}\|_{\infty}^M$ respectively scale as $\sim Re^{1.1}$ for plane Couette flow and $\sim Re^{1.5}$ for plane Poiseuille flow, which matches the scaling of $\|\mathcal{H}_{\nabla}\|_{\mu}$. On the other hand, $\|\mathcal{H}_{\nabla vy}\|_{\infty}^M$ is much smaller  \rev{than these three quantities and} scales as $\|\mathcal{H}_{\nabla vy}\|_{\infty}^M\sim Re$ for both plane Couette and Poiseuille flows.

\rev{In order to understand the role of the oblique waves in this scaling, we also   plot the quantity $\|\mathcal{H}_{\nabla}\|_{\mu}(1,1)$ corresponding to the wavenumber pair associated with the oblique waves discussed in \S\ \ref{subsec:scale_dependent_mu_cou} and \S\ \ref{subsec:scale_dependent_mu_poi} in figure \ref{fig:Re_dependence}. In both flows, these values are lower than and not exactly parallel with $\|\mathcal{H}_{\nabla}\|_{\mu}^M$, indicating that they are not associated with the peak amplification of $\|\mathcal{H}_{\nabla}\|_{\mu}$ in figure \ref{fig:mu_cou}(a) and \ref{fig:mu_poi}(a). However, they do seem to provide the majority of the contribution to $\|\mathcal{H}_{\nabla}\|_{\mu}^M$. This observation is consistent with figures \ref{fig:mu_cou}(a) and \ref{fig:mu_poi}(a), where $(k_x,k_z)=(1,1)$ is close to but different from the $(k_x^M,k_z^M)$ wavenumber pair that reaches the maximum value of $\|\mathcal{H}_{\nabla}\|_{\mu}(k_x,k_z)$ defined as:
\begin{align}
    (k_x^M, k_z^M):=\underset{k_x,k_z}{\text{arg max }}\|\mathcal{H}_{\nabla}\|_{\mu}(k_x,k_z). 
\end{align}
These wavenumber pairs are $(k_x^M,k_z^M)=(0.19, 0.58)$ for plane Couette flow at $Re=358$ and $(k_x^M, k_z^M)=(0.69,1.56)$ for plane Poiseuille flow at $Re=690$. We also plot the Reynolds number dependence of $\|\mathcal{H}_{\nabla}\|_{\mu}(0.19,0.58)\sim Re$ for plane Couette flow and $\|\mathcal{H}_{\nabla}\|_{\mu}(0.69,1.56)\sim Re^{1.3}$ for plane Poiseuille flow as ({\color{red}$\times \dashed$ }, red) markers in figure \ref{fig:Re_dependence}. Here, we observe that they overlap with  $\|\mathcal{H}_{\nabla}\|_{\mu}^M$ at low Reynolds numbers, but deviate as the Reynolds number increases leading to a reduced $Re$ scaling compared with $\|\mathcal{H}_{\nabla}\|_{\mu}^M$. These results show that these oblique flow structures continue to show very high (close to the maximum overall amplification value) throughout the Reynolds number range. However, the wavenumber pair $(k_x^M, k_z^M)$ that reaches the maximum value over $\|\mathcal{H}_{\nabla}\|_{\mu}(k_x,k_z)$  depends on the Reynolds number. }

The observed importance \rev{and analytical tractability} of streamwise elongated structures have motivated previous analysis of the streamwise constant $(k_x=0)$ component of the frequency response operator. \rev{In order to compare our analysis to these results, we also} evaluate this behavior by computing analogous quantities 
\begin{subequations}
\begin{align}
    \|\mathcal{H}_{\nabla }\|_{\mu}^{sc}:=&\underset{k_z,\,k_x=10^{-4}}{\text{max}}\|\mathcal{H}_{\nabla }\|_{\mu}(k_x,k_z), \\
    \|\mathcal{H}_{\nabla ij}\|_{\infty}^{sc}:=&\underset{k_z,\,k_x=10^{-4}}{\text{max}}\|\mathcal{H}_{\nabla ij}\|_{\infty}(k_x,k_z), 
    \label{eq:H_nabla_ij_M_sc}
\end{align}
\end{subequations}
which restricts the streamwise wavenumber to $k_x=10^{-4}$ to approximate the streamwise constant modes. In figure \ref{fig:Re_dependence_isolated_kx0}, we replot $\|\mathcal{H}_{\nabla }\|_{\mu}^{M}$ alongside $\|\mathcal{H}_{\nabla }\|_{\mu}^{sc}$ ($\triangle\dashed$, black) , and observe that $\|\mathcal{H}_{\nabla }\|_{\mu}^{sc}\sim Re$ for both plane Couette and Poiseuille flows. Figure \ref{fig:Re_dependence_isolated_kx0} also shows  $\|\mathcal{H}_{\nabla ux}\|_{\infty}^{sc}$, $\|\mathcal{H}_{\nabla vy}\|_{\infty}^{sc}$, and $\|\mathcal{H}_{\nabla wz}\|_{\infty}^{sc}$. Here, we find that $\|\mathcal{H}_{\nabla }\|_{\mu}^{sc}$ overlaps with $\|\mathcal{H}_{\nabla ux}\|_{\infty}^{sc}$ and thus shows the same scaling $\sim Re$. These three input--output pathways $\|\mathcal{H}_{\nabla ij}\|_{\infty}^{sc}$ ($ij=ux,\;vy,\;wz$) scale as $\sim Re$ for both plane Couette and Poiseuille flows. This behavior is consistent with the results in \citet[theorem 11]{jovanovic2004modeling}, which showed that $\|\mathcal{H}_{ij}\|_{\infty}\sim Re$ ($ij=ux,\;vy,\;wz$) when it is restricted to $k_x=0$ for both plane Couette and plane Poiseuille flows. \rev{The following theorem provides an analogous analytical $Re$ scaling for $\|\mathcal{H}_{\nabla ij}\|_{\infty}$ at  $k_x=0$, i.e., the streamwise constant component.} 
\rev{

\begingroup
\allowdisplaybreaks
\begin{thm}
\label{thm:scaling_Re_Pr}
Given streamwise constant ($k_x=0$) plane Couette flow or plane Poiseuille flow. Each component of $\|\mathcal{H}_{\nabla ij}\|_{\infty}$ ($i=u,v,w$ and $j=x,y,z$) scales as:
\begin{align}
    & \begin{bmatrix}
    \|\mathcal{H}_{\nabla ux}\|_{\infty} & \|\mathcal{H}_{\nabla uy}\|_{\infty} & \|\mathcal{H}_{\nabla uz}\|_{\infty} \\
    \|\mathcal{H}_{ \nabla vx}\|_{\infty} & \|\mathcal{H}_{\nabla vy}\|_{\infty} & \|\mathcal{H}_{\nabla vz}\|_{\infty}\\
    \|\mathcal{H}_{\nabla wx}\|_{\infty} & \|\mathcal{H}_{\nabla wy}\|_{\infty} & \|\mathcal{H}_{\nabla wz}\|_{\infty}\\
\end{bmatrix} \nonumber \\
=&\begin{bmatrix}
    Re\, h_{\nabla ux}(k_z) & Re^2\, h_{\nabla uy}(k_z) & Re^2\, h_{ \nabla uz}(k_z) \\
    0 & Re\,h_{\nabla vy}(k_z) & Re\,h_{ \nabla vz}( k_z) \\
    0 & Re\, h_{\nabla wy}(k_z) & Re \,h_{ \nabla wz}(k_z)
    \end{bmatrix},
\end{align}
where functions $h_{\nabla ij}(k_z)$ are independent of the $Re$. 
\end{thm}

\endgroup
}
\rev{The proof of theorem \ref{thm:scaling_Re_Pr} in Appendix \ref{appendix:proof_scaling_Re_Pr} follows the procedure in \citep{jovanovic2004modeling,Jovanovic2005,jovanovic2020bypass}, which involves the change of variable $\Omega:=\omega Re$. Comparing scaling of this $\|\mathcal{H}_{\nabla ij}\|_{\infty}$ in theorem \ref{thm:scaling_Re_Pr} with that of $\|\mathcal{H}_{ij}\|_{\infty}$ at $k_x=0$ in \citet[theorem 11]{jovanovic2004modeling} shows that the modification of the operator to provide the output $\widehat{\boldsymbol{\nabla}}\widehat{u}$, $\widehat{\boldsymbol{\nabla}}\widehat{v}$, and $\widehat{\boldsymbol{\nabla}}\widehat{w}$ does not modify the Reynolds number scaling, which is expected since this operation amounts to a Reynolds number independent transformation of the system output.}

\rev{
Combining results in theorems \ref{lemma:mu_componentwise_inf}-\ref{thm:scaling_Re_Pr}, we have the following corollary \ref{thm:scaling_mu} providing an analytical lower bound of $\|\mathcal{H}_{\nabla}\|_{\mu}$ at $k_x=0$. 
\begin{corollary}
Given streamwise constant ($k_x=0$) plane Couette flow or plane Poiseuille flow.
\begin{align}
    \|\mathcal{H}_{\nabla}\|_{\mu}(0,k_z)\geq \text{max}[Re\, h_{\nabla ux }(k_z), Re\, h_{\nabla vy }(k_z),Re\, h_{\nabla wz }(k_z)],
    \label{eq:mu_componentwise_inequality_Re_Pr}
\end{align}
where functions $h_{\nabla ij}(k_z)$ with $ij=ux,vy,wz$ are independent of the $Re$. 
\label{thm:scaling_mu}
\end{corollary}
The previous numerical observations of $\|\mathcal{H}_\nabla\|_\mu^{sc}\sim Re$ in figure \ref{fig:Re_dependence_isolated_kx0} for both plane Couette and plane Poiseuille flows are consistent with corollary \ref{thm:scaling_mu}. 
}

The reduced scaling exponent $\eta$ of the largest structured gain $\|\mathcal{H}_{\nabla}\|_{\mu}^M\sim Re^\eta$ observed here compared with $\eta=2$ for unstructured gain \citep{trefethen1993hydrodynamic,kreiss1994bounds,jovanovic2004modeling} further highlights the importance of the componentwise structure of nonlinearity imposed in this framework, which appears to weaken the large amplification associated with the lift-up mechanism.

\section{Conclusions and future work}
\label{sec:conclusion}

This work proposes a \emph{structured} input--output analysis that augments the traditional spatio-temporal frequency response with structured uncertainty. The structure preserves the componentwise input--output structure of the nonlinearity in \rev{the NS} equations. We then analyze the spatio-temporal response of the resulting feedback interconnection between the \rev{LNS equations} and the structured forcing  in terms of the structured singular value of the associated spatio-temporal frequency response operator.

We apply the structured input--output analysis to transitional plane Couette and plane Poiseuille flows. Comparisons of the results to those of traditional analysis and an unstructured feedback interconnection indicate that the addition of a structured feedback interconnection enables the prediction of a wider range of known dominant flow structures to be identified without the computational burden of nonlinear optimization or extensive simulations. More specifically, the results for transitional plane Couette flow reproduce the findings from direct numerical simulation (DNS) \citep{reddy1998stability} and nonlinear optimal perturbation (NLOP) \citep{rabin2012triggering} in showing that oblique waves require less energy to \rev{induce transition} than the streamwise elongated structures emphasized in traditional input--output analysis. In plane Poiseuille flow the results again predict the oblique wave structure as in DNS \citep{reddy1998stability}. They also highlight the importance of spatially localized structures with a streamwise wavelength larger than spanwise similar to NLOP~\citep{farano2015hairpin}. The framework also reproduces the oblique turbulent bands \citep{prigent2003long,kanazawa2018lifetime} that have been associated with transitioning flows with very large channel sizes  ($\sim O(100)$ times the channel half-height) in both experiments and DNS.

The agreement between the predictions from structured input--output analysis and observation in experiments, DNS, and NLOP indicate that the structured feedback interconnection reproduces  important nonlinear effects. Our analysis suggests that restricting the feedback pathways preserves the structure of the nonlinear mechanisms that weaken the streaks developed through the lift-up effect, in which cross-stream forcing amplifies streamwise streaks \citep{ellingsen1975stability,landahl1975wave,Brandt2014}. \rev{Traditional input--output analysis instead predicts the dominance of streamwise elongated structures associated with the lift-up mechanism, see e.g. the discussion in \citet{jovanovic2020bypass}.} The Reynolds number dependence observed in our studies further \rev{supports the notion that imposing a structured feedback interconnection based on certain input--output properties associated with the nonlinearity in the NS equations leads to a weakening of the amplification of streamwise elongated structures.}



The results here suggest the promise of this computationally tractable approach and opens up  many directions for future work. \rev{Further refinement of the structured uncertainty may provide additional physical insight. This extension and the development of the associated computational tools are the subjects of ongoing work. The results here are associated with the maximum amplification over all frequencies but it may be also interesting to isolate each temporal frequency and examine the frequency that maximizes the amplification under this structured feedback interconnection.
} \rev{Another} natural direction is an extension to pipe flow, where the subcritical transition is also widely studied; see e.g., \citep{hof2003scaling,peixinho2007finite,eckhardt2007turbulence,mellibovsky2009critical,mullin2011experimental,Pringle2010,Pringle2012,barkley2016theoretical}. Adaptions of this approach to the fully developed turbulent regime, where the resolvent framework and input--output analysis have provided important insights is another direction of ongoing study.

\section*{Acknowledgment}
The authors gratefully acknowledge support from the US National Science Foundation (NSF) through grant number CBET 1652244 and the Office of Naval Research (ONR) through grant number N00014-18-1-2534. C.L. greatly appreciates the support from the Chinese Scholarship Council.

\section*{Declaration of Interests}
The authors report no conflict of interest.

\rev{
\appendix

\iftoggle{thesis}{\chapter{Proof of theorems \ref{lemma:mu_componentwise_inf}-\ref{thm:scaling_Re_Pr}} }{\section{Proof of theorems \ref{lemma:mu_componentwise_inf}-\ref{thm:scaling_Re_Pr}} }

\iftoggle{thesis}{\section{Proof of theorem \ref{lemma:mu_componentwise_inf}}}{\subsection{Proof of theorem \ref{lemma:mu_componentwise_inf}}}
\label{appendix:mu_componentwise_inf}

\begingroup
\allowdisplaybreaks
\begin{myproof}
We define the following sets of uncertainty:
\begin{subequations}
\label{eq:uncertain_set_appendix}
\begin{align}
\mathbfsbilow{\widehat{U}}_{\Upxi,ux}:=\left\{\text{diag}\left(-\mathbfsbilow{\widehat{u}}_{\xi}^{\text{T}},\mathsfbi{0},\mathsfbi{0}\right):-\mathbfsbilow{\widehat{u}}^{\text{T}}_{\xi}\in \mathbb{C}^{N_y\times 3N_y}\right\},\label{eq:uncertain_set_u}\\
\mathbfsbilow{\widehat{U}}_{\Upxi,vy}:=\left\{\text{diag}\left(\mathsfbi{0},-\mathbfsbilow{\widehat{u}}_{\xi}^{\text{T}},\mathsfbi{0}\right):-\mathbfsbilow{\widehat{u}}^{\text{T}}_{\xi}\in \mathbb{C}^{N_y\times 3N_y}\right\},\label{eq:uncertain_set_v}\\
\mathbfsbilow{\widehat{U}}_{\Upxi,wz}:=\left\{\text{diag}\left(\mathsfbi{0},\mathsfbi{0},-\mathbfsbilow{\widehat{u}}_{\xi}^{\text{T}}\right):-\mathbfsbilow{\widehat{u}}^{\text{T}}_{\xi}\in \mathbb{C}^{N_y\times 3N_y}\right\}.\label{eq:uncertain_set_w}
\end{align}
\end{subequations}
Here, $\mathsfbi{0}\in \mathbb{C}^{N_y\times 3N_y}$ is a zero matrix with the size $N_y\times 3N_y$. 
Then, using the definition of the structured singular value in definition \ref{def:mu}, we have:
\begin{subequations}
\label{eq:mu_componentwise}
\begin{align}
    &\mu_{\mathbfsbilow{\widehat{U}}_{\Upxi,ux}}\left[\mathbfsbilow{H}_{\nabla}(k_x,k_z,\omega)\right]\nonumber\\
    =&\frac{1}{\text{min}\{\bar{\sigma}[\mathbfsbilow{\widehat{u}}_{\Upxi,ux}]\,:\,\mathbfsbilow{\widehat{u}}_{\Upxi,ux}\in \mathbfsbilow{\widehat{U}}_{\Upxi,ux},\,\text{det}[\mathsfbi{I}-\mathbfsbilow{H}_{\nabla}(k_x,k_z,\omega)\mathbfsbilow{\widehat{u}}_{\Upxi,ux}]=0\}}\label{eq:mu_componentwise_1}\\
    =&\frac{1}{\text{min}\{\bar{\sigma}[-\mathbfsbilow{\widehat{u}}^{\text{T}}_{\xi}]\,:\,-\mathbfsbilow{\widehat{u}}^{\text{T}}_{\xi}\in \mathbb{C}^{N_y\times 3N_y},\,\text{det}[\mathsfbi{I}_{3N_y}-\mathbfsbilow{H}_{\nabla ux}(k_x,k_z,\omega)(-\mathbfsbilow{\widehat{u}}^{\text{T}}_{\xi})]=0\}}\label{eq:mu_componentwise_2}\\
    =&\bar{\sigma}[\mathbfsbilow{H}_{\nabla ux}(k_x,k_z,\omega)].\label{eq:mu_componentwise_3}
\end{align}
\end{subequations}
Here, the equality \eqref{eq:mu_componentwise_1} is obtained by substituting the uncertainty set in \eqref{eq:uncertain_set_u} into definition \ref{def:mu}. The equality \eqref{eq:mu_componentwise_2} is obtained by performing block diagonal partition of terms inside of $\text{det}[\cdot]$ and employing zeros in the uncertainty set in equation \eqref{eq:uncertain_set_u}. Here, $\mathbfsbilow{H}_{\nabla ux}$ is the discretization of $\mathcal{H}_{\nabla ux}$ and $\mathsfbi{I}_{3N_y}\in \mathbb{C}^{3N_y\times 3N_y} $ in \eqref{eq:mu_componentwise_2} is an identity matrix with matching size $(3N_y\times 3N_y)$, whereas  $\mathsfbi{I} \in \mathbb{C}^{9N_y\times 9N_y}$ in \eqref{eq:mu_componentwise_1}. The equality \eqref{eq:mu_componentwise_3} is using the definition of unstructured singular value; see e.g., \citep[equation (11.1)]{zhou1996robust}. 

Similarly, we have:
\begin{subequations}
\label{eq:mu_componentwise_vwrho}
\begin{align}
    \mu_{\mathbfsbilow{\widehat{U}}_{\Upxi,vy}}\left[\mathbfsbilow{H}_{\nabla}(k_x,k_z,\omega)\right]=\bar{\sigma}[\mathbfsbilow{H}_{\nabla vy}(k_x,k_z,\omega)],\\
    \mu_{\mathbfsbilow{\widehat{U}}_{\Upxi,wz}}\left[\mathbfsbilow{H}_{\nabla}(k_x,k_z,\omega)\right]=\bar{\sigma}[\mathbfsbilow{H}_{\nabla wz}(k_x,k_z,\omega)].
\end{align}
\end{subequations}
Using the fact that $\mathbfsbilow{\widehat{U}}_{\Upxi}\supseteq \mathbfsbilow{\widehat{U}}_{\Upxi,ij}$ with $ij=ux,vy,wz$ and equalities in \eqref{eq:mu_componentwise}-\eqref{eq:mu_componentwise_vwrho}, we have:
\begin{align}
     \mu_{\mathbfsbilow{\widehat{U}}_{\Upxi}}\left[\mathbfsbilow{H}_{\nabla}(k_x,k_z,\omega)\right]\geq\mu_{\mathbfsbilow{\widehat{U}}_{\Upxi,ij}}\left[\mathbfsbilow{H}_{\nabla}(k_x,k_z,\omega)\right]=\bar{\sigma}[\mathbfsbilow{H}_{\nabla ij}(k_x,k_z,\omega)].
     \label{eq:mu_componentwise_inequality}
\end{align}
Applying the supreme operation $\underset{\omega \in \mathbb{R}}{\text{sup}}[\cdot]$ on \eqref{eq:mu_componentwise_inequality} and using definitions of $\|\cdot\|_{\mu}$ and $\|\cdot \|_{\infty}$ we have:
\begin{subequations}
\begin{align}
    \|\mathcal{H}_{\nabla }\|_{\mu}\geq \|\mathcal{H}_{\nabla ux}\|_{\infty},\;\; \|\mathcal{H}_{\nabla }\|_{\mu}\geq \|\mathcal{H}_{\nabla vy}\|_{\infty},\;\;
    \|\mathcal{H}_{\nabla }\|_{\mu}\geq \|\mathcal{H}_{\nabla wz}\|_{\infty}.\tag{\theequation a-c}
\end{align}
\end{subequations}
This directly results in inequality \eqref{eq:mu_larger_than_all_diagonal_component} of theorem \ref{lemma:mu_componentwise_inf}. 
\end{myproof}
\endgroup

\iftoggle{thesis}{\section{Proof of theorem \ref{thm:scaling_Re_Pr}}}{\subsection{Proof of theorem \ref{thm:scaling_Re_Pr}}}

\label{appendix:proof_scaling_Re_Pr}

\begin{myproof}
Under the assumptions of streamwise constant $k_x=0$ for plane Couette flow or plane Poiseuille flow in theorem \ref{thm:scaling_Re_Pr}, the operator $\widehat{\mathcal{A}}$, $\widehat{\mathcal{B}}$, and $\widehat{\mathcal{C}}$ can be simplified and decomposed as:
\begin{subequations}
\label{eq:operator_ABC_appendix_proof}
\begin{align}
    \widehat{\mathcal{A}}(k_x,k_z)=&\begin{bmatrix}
   \frac{\widehat{{\nabla}}^{-2}\widehat{{\nabla}}^4}{Re} & 0 \\
    -\text{i}k_zU' & \frac{\widehat{{\nabla}}^2}{Re} 
    \end{bmatrix},\\
    \mathcal{\widehat{B}}(k_x,k_z)=&
    \begin{bmatrix}
    0 & -k_z^2\widehat{{\nabla}}^{-2} & -\text{i}k_z\widehat{{\nabla}}^{-2} \partial_y \\
    \text{i}k_z & 0 & 0 
    \end{bmatrix}=:\begin{bmatrix}
    0 & \widehat{\mathcal{B}}_{y,1} & \widehat{\mathcal{B}}_{z,1} \\
    \mathcal{B}_{x,2} & 0 & 0
    \end{bmatrix},\\
    \mathcal{\widehat{C}}(k_x,k_z)=&\begin{bmatrix}
    0 & -\text{i}/k_z  \\
    \mathcal{I} & 0 \\
    \text{i} \partial_y/k_z & 0
    \end{bmatrix}=:\begin{bmatrix}
    0 &  \mathcal{\widehat{C}}_{u,2}  \\
     \mathcal{\widehat{C}}_{v,1} & 0 \\
     \mathcal{\widehat{C}}_{w,1} & 0 
    \end{bmatrix}.
\end{align}
\end{subequations}
Here, we employ a matrix inverse formula for the lower triangle block matrix:
\begin{align}
        \begin{bmatrix}
    L_{11} & 0 \\
    L_{21} & L_{22}
    \end{bmatrix}^{-1}=\begin{bmatrix}
    L_{11}^{-1} & 0\\
    -L_{22}^{-1}L_{21}L_{11}^{-1} & L_{22}^{-1} 
    \end{bmatrix}
\end{align}
to compute $\left(\text{i}\omega\mathcal{I}_{2\times 2}-\widehat{\mathcal{A}}\right)^{-1}$. Then, we employ a change of variable $\Omega=\omega Re$ similar to \citep{jovanovic2004modeling,Jovanovic2005,jovanovic2020bypass} to obtain $\mathcal{H}_{\nabla ij}$ with $i=u,v,w$, and $j=x,y,z$ as:
\begingroup
\allowdisplaybreaks
\begin{subequations}
\begin{align}
    \mathcal{H}_{\nabla ux}=&\widehat{\boldsymbol{\nabla}}\widehat{\mathcal{C}}_{u,2}Re\left(\text{i}\Omega \mathcal{I}-\widehat{{\nabla}}^2\right)^{-1}\widehat{\mathcal{B}}_{x,2},\\
    \mathcal{H}_{\nabla uy}=&\widehat{\boldsymbol{\nabla}}\widehat{\mathcal{C}}_{u,2}Re\left(\text{i}\Omega \mathcal{I}-\widehat{{\nabla}}^2\right)^{-1}(-\text{i}k_zU')Re\left(\text{i}\Omega \mathcal{I}-\widehat{{\nabla}}^{-2}\widehat{{\nabla}}^{4}\right)^{-1} \widehat{\mathcal{B}}_{y,1},\\
    \mathcal{H}_{\nabla uz}=&\widehat{\boldsymbol{\nabla}}\widehat{\mathcal{C}}_{u,2}Re\left(\text{i}\Omega \mathcal{I}-\widehat{{\nabla}}^2\right)^{-1}(-\text{i}k_zU')Re\left(\text{i}\Omega \mathcal{I}-\widehat{{\nabla}}^{-2}\widehat{{\nabla}}^{4}\right)^{-1} \widehat{\mathcal{B}}_{z,1},\\
    \mathcal{H}_{\nabla vx}=&0,\\
    \mathcal{H}_{\nabla vy}=&\widehat{\boldsymbol{\nabla}}\widehat{\mathcal{C}}_{v,1}Re\left(\text{i}\Omega \mathcal{I}-\widehat{{\nabla}}^{-2}\widehat{{\nabla}}^{4}\right)^{-1} \widehat{\mathcal{B}}_{y,1},\\
    \mathcal{H}_{\nabla vz}=&\widehat{\boldsymbol{\nabla}}\widehat{\mathcal{C}}_{v,1}Re\left(\text{i}\Omega \mathcal{I}-\widehat{{\nabla}}^{-2}\widehat{{\nabla}}^{4}\right)^{-1} \widehat{\mathcal{B}}_{z,1},\\ \mathcal{H}_{\nabla wx}=&0,\\
    \mathcal{H}_{\nabla wy}=&\widehat{\boldsymbol{\nabla}}\widehat{\mathcal{C}}_{w,1}Re\left(\text{i}\Omega \mathcal{I}-\widehat{{\nabla}}^{-2}\widehat{{\nabla}}^{4}\right)^{-1} \widehat{\mathcal{B}}_{y,1},\\
    \mathcal{H}_{\nabla wz}=&\widehat{\boldsymbol{\nabla}}\widehat{\mathcal{C}}_{w,1}Re\left(\text{i}\Omega \mathcal{I}-\widehat{{\nabla}}^{-2}\widehat{{\nabla}}^{4}\right)^{-1} \widehat{\mathcal{B}}_{z,1}. 
\end{align}
\end{subequations}
\endgroup
Taking the operation that $\|\cdot\|_{\infty}=\underset{\omega\in\mathbb{R}}{\text{sup}}\bar{\sigma}[\cdot]=\underset{\Omega\in\mathbb{R}}{\text{sup}}\bar{\sigma}[\cdot]$, we have the scaling relation in theorem \ref{thm:scaling_Re_Pr}. 
\end{myproof}
}
\bibliography{main}
\bibliographystyle{jfm}

\end{document}